\documentclass[english]{revtex4-2}
\usepackage[T1]{fontenc}
\usepackage[latin9]{inputenc}
\usepackage{color}
\usepackage{float}
\usepackage{amstext}
\usepackage{amssymb}
\usepackage{graphicx}
\usepackage{babel}
\begin{document}
\title{Robust applicability of continuous dynamical decoupling to decoherence
reduction in longitudinal and transverse-noise settings: The role
of anisotropy}
\author{S. Afonso, J.M. Gomez Llorente, and J. Plata}
\address{Departamento de Física and IUdEA, Universidad de La Laguna,~\\
 La Laguna E38200, Tenerife, Spain.}
\begin{abstract}
We analytically evaluate the efficiency of continuous dynamical decoupling
(CDD) to curb decoherence in generic qubit setups where diverse sources
of noise can be present. Previous theoretical approaches to CDD have
mainly focused on its potential to cope with \emph{longitudinal} fluctuations.
Here, the basic scenario tackled with CDD is generalized. Apart from
dealing with pure dephasing induced by \emph{diagonal} noise, we consider
the impact of \emph{transverse} fluctuations, usually present in the
practical arrangements. In particular, the implications of anisotropic
noisy inputs are studied. Additionally, we analyze the role of the
fluctuations in the dressing of the qubit by the CDD field of control:
since the driving field is usually switched on through linear ramps
of its characteristic parameters, the associated dressing of the original
states can be described in terms of noisy Landau-Zener transitions.
In our approach, based on a sequence of unitary transformations, the
noise entering the system is cast into effective stochastic terms
whose spectral characteristics are dependent on the driving parameters.
This description allows the design of strategies  to mitigate the
impact of the fluctuations using controlled changes in the effective-noise
properties. Significant robustness of CDD against the generalization
of the basic scenario can be achieved through an appropriate choice
of the parameters of control. 
\end{abstract}
\maketitle

\section{Introduction}

Quantum technologies are playing a leading role in the fast development
of fields like information processing, metrology, or sensing \citep{key-Preskill,key-QuantumSensingRMP,key-AcinetalReviewQuantumTech,key-SuterAlvarezRMPdecoherenceGeneral,key-Plenio2Sensing,key-MonroeTrappedIonsEngineering,key-Metrology}.
As they are based on resources specifically associated to the quantum
character of the dynamics, their potential depends crucially on maintaining
the intrinsic quantum features in the evolution of the applied systems.
Indeed, a central objective of the research in these fields is the
realization of technical schemes that allow controlling the dynamics
while preserving the quantum characteristics. Particularly important
in this context is the decoherence problem: the loss of \emph{purity}
generated by the coupling of the primary system to non-controllable
environments is a fundamental difficulty in the realization of intrinsically
quantum effects. For instance, the relevance of the \emph{decoherence
issues} to the feasibility of the quantum-computing protocols is decisive:
the power associated to the possibility of working with a superposition
of states disappears when the information on the relative phases is
lost, i.e., when decoherence sets in. Hence, it is understood that
the implementation of methods for curbing the effect of environment-induced
fluctuations, and, consequently, for extending the coherence times,
is a basic requirement for the applicability of the designed protocols.
Apart from technical importance, preserving the coherence has central
relevance to fundamental areas of research. In this sense, it is worth
stressing the crucial role that it has played in the realization of
a variety of novel  effects with ultracold atoms (see \citep{key-Spielman2,key-Spielman3SimulNonAbelianMonopole,key-BlochReview,key-MagneticShield}
and references therein).

A basic strategy in the design of methods for decoherence reduction
consists in effectively disconnecting the system from the environment
that generates the fluctuations. That is the basis of the techniques
of dynamical decoupling \citep{key-HahnDDorigins,key-ViolaDDpulsesOrigins,key-Viola2RobustDD,key-UhrigDDpulsesBasic}.
In order to achieve the effective decoupling two main schemes have
been applied. The first one consists in using sequences of pulses
of control to average out the effect of the fluctuations \citep{key-AlvarezDDpulseComparingSchemes,key-ItanoDDpulseOptimSequences,key-LidarDDpulseDesign}.
The second one, on which we focus here, the so called continuous dynamical
decoupling (CDD) method, employs, instead of pulses, continuous-wave
driving fields intended to facilitate the integration of the information
protocols, and to simplify the experimental realization \citep{key-MiaoAwschalomCDDSolidStateSpinQubit,key-FanchiniCDDOptimFieldDesign,key-PlenioConcatenationTrappedIons,key-Plenio3ions}.
In its basic version, the method incorporates a driving field quasiresonant
and \emph{orthogonal} to the qubit, i.e., in a direction perpendicular
to the qubit quantization axis. The aim is to relegate the random
components to a secondary role in the (driven) dynamics. Indeed, for
a sufficiently large driving amplitude, a qubit built from the dressed
states is guaranteed to be significantly protected from noise. Additionally,
to deal with the fluctuations introduced into the system via the driving-field
intensity, a concatenation scheme is incorporated: a second field
of control, orthogonal to the first one and with characteristic parameters
conveniently chosen, is applied to cope with the extra noise \citep{key-PlenioConcatenationTrappedIons}.
Originally aimed at dealing with the effect of static noise entering
diagonally the qubits, the CDD techniques have found applicability
in more general settings. Recent work \citep{key-GomezLlorenteNoiseSpectrum}
on the dependence of the method performance on the noise-spectrum
characteristics provided the clues to understanding the observed applicability
of CDD beyond the static-noise scenario. Moreover, the relevance of
the technique to effective non-Hermitian Hamiltonians, operatively
accounting for dissipation, has recently been tested \citep{key-CDDNonHermitian}.
Also noticeable are variations on the method implementation like the
use of phase shifts in the driving field as an additional tool of
control \citep{key-ContinuousPhaseDecoupling,key-BarGillPhaseModulatedExpReal},
the combination of a continuous field with interrogation pulses \citep{key-QuasiContinuousIon}
or with sequences of phased pulses \citep{key-MixedDynamicalDecoupling},
the design of protocols for field optimization \citep{key-OptimizationMachineLearning,key-OptimizationGeometry},
or the controlled use of destructive interference of noise sources
\citep{key-DestructiveInterference}. Substantial advances in the
performance of the method have been achieved in contexts like trapped
ions \citep{key-MultiIonFrequency,key-BermudezTrappedIons,key-PlenioConcatenationTrappedIons,key-Plenio3ions,key-TrappedIonsQuadrupole},
solid-state spin qubits \citep{key-SolidStatSpinQubitProtect,key-MiaoAwschalomCDDSolidStateSpinQubit},
superconducting circuits \citep{key-CDDsuperconductingTransmon},
NV centers in diamond \citep{key-XuDiamond}, clock transitions in
atomic systems \citep{key-SpielmanClockTransitions,key-StarkClock3levelsNVcenters},
or microwave quantum heterodyne sensing \citep{key-HeterodyneSensing}.
Here, to give further theoretical support to the applicability of
CDD, we will  extend the basic model along two lines. First, the effect
of non-diagonal fluctuations on the method efficiency will be assessed.
Actually, the characterization of the role of transverse noise in
decoherence is an open problem which is being the subject of intense
research. Particular interest exists in tracing the implications of
anisotropy \citep{key-AnisotropicFaultTolerant,key-AnisotropyBosco,key-AnisotropyChen,key-AnisotropyChoiJoynt,key-AnisotropyHayati,key-AnisotropyHendrickx,key-AnisotropySaezMollejo}.
In this context, it is worth referring to recent work on a \emph{fluxonium}
qubit in a non-diagonal-noise setting \citep{key-AnisotropicTransverseNoise},
where anisotropy was found to lead to nontrivial evolution of the\emph{
purity}. There, to facilitate the systematic analysis of the problem,
a controlled scenario was implemented with designed random signals
emulating different forms of anisotropic noise. Our study can be interpreted
as corresponding to a (generalized) driven version of that model.
In a second line, we will deal with an inevitable hindrance of the
CDD implementation, namely, the presence of noise during the practical
preparation of the dressed states. The importance of this issue is
evident: the lack of precision in the generation of the dressed states
can lead to errors in the subsequent performance of the CDD qubit.
In both lines, no restrictions on the magnitude of the noise correlation
time will be assumed. Indeed, our general description incorporates
the white-noise limit and the static-noise regime as particular cases.
In our approach, based on an operative picture of the dressed-state
framework obtained via a sequence of time-dependent unitary transformations,
the fluctuations entering the system are cast into effective random
terms whose spectral properties are dependent on the field characteristics.
This scheme allows considering an appropriate choice of the driving
parameters as a strategy to control the effective spectral densities,
and, in turn, to reduce the effect of noise. With the obtained results,
the applicability and efficiency of the CDD technique can be evaluated
for various experimental conditions. In particular, our methodology
will be illustrated by its application to the setup of Ref. \citep{key-SpielmanClockTransitions},
used in the realization of clock transitions in atomic systems. That
scenario exemplifies the potential of the method to curb decoherence
and provides us with a prototype  system where the components and
requirements for the CDD to be operative can be appropriately checked.
As that realization involves a (generic) hyperfine Zeeman multiplet
with no restrictions on the quantum number $F$, it can be regarded
as a qudit setup. Magnetic noise present in this scenario, will be
modelled as a generic Ornstein-Uhlenbeck (OU) process. The limits
of static fluctuations and white noise will be explicitly tackled.

The outline of the paper is as follows. In Sec. II, the theoretical
basis of the CDD technique will be generalized by including both longitudinal
and transverse fluctuations in the Hamiltonian that governs the qubit
dynamics. A sequence of unitary transformations will be applied to
provide a compact description of  the dressed-state qubit. Additionally,
the properties of the effective noise terms that emerge from the original
fluctuations via the unitary transformations will be characterized.
In Sec. III, the conditions that guarantee the sound applicability
of the method will be identified: the requirements for reducing dephasing
and stochastic population transfer will be evaluated. As demanded
for the discussion of some of the obtained results, the description
will be taken beyond the Rotating Wave Approximation (RWA). Some aspects
of this generalized picture are presented in Appendix A. Moreover,
some technical details of our procedure are given in Appendix B. Additionally,
in Appendix C, we extend our scheme to incorporate (transverse) amplitude
noise, i.e., fluctuations in the amplitude of the driving field. In
Sec. IV, we present a numerical simulation of the decoherence process.
The predictions of our theoretical approach will be confirmed by the
numerical results. Sec. V will be focused on the analytical study
of the effect of the fluctuations on the preparation of the dressed
states. We will assess the robustness against noise of the techniques
of adiabatic passage implemented via linear ramps of the amplitude
and detuning of the driving field. To this end, the Landau-Zener model
\citep{key-Landau,key-Zener} will be generalized to incorporate the
fluctuations. Finally, some general conclusions will be summarized
in Sec. VI. 

\section{Generalization of the CDD scenario}

\subsection{The CDD Hamiltonian with longitudinal and transverse fluctuations}

The generalization of the CDD framework will be built from the model-system
corresponding to the realization of clock transitions of Ref. \citep{key-SpielmanClockTransitions}.
There, CDD was applied to  curb the effect of magnetic noise on a
hyperfine Zeeman multiplet in $^{87}\textrm{Rb}$. Accordingly, we
deal with the Hamiltonian

\begin{eqnarray}
\hat{H} & = & \left[\omega_{0}+\delta\omega_{0}(t)\right]\hat{F}_{z}+2\Omega_{d}\cos(\omega_{d}t)\hat{F}_{x}+\nonumber \\
 &  & \eta_{x}(t)\hat{F}_{x}+\eta_{y}(t)\hat{F}_{y}\label{eq:BasicHamiltonian}
\end{eqnarray}
 where $\mathbf{\hat{F}}$ is the angular momentum operator, (the
states of the associated representation will be denoted as $\left|F,m\right\rangle $).
$\omega_{0}$ stands for the mean value of the multiplet frequency.
(In our model, we do not exclusively refer to $^{87}\textrm{Rb}$;
without loss of generality, we will consider $\omega_{0}>0$). $\Omega_{d}$
and $\omega_{d}$ are parameters characteristic of the control field:
$\Omega_{d}$ is proportional to the field amplitude and to the Landé
factor of the multiplet, and, $\omega_{d}$ denotes the field frequency.
The model also includes longitudinal and transverse fluctuations:
$\delta\omega_{0}(t)$ is the frequency shift induced by the (longitudinal)
fluctuations in the bias magnetic field that splits the multiplet,
and $\eta_{x}(t)$ and $\eta_{y}(t)$ account for transverse noise,
potentially present in the experimental setups. The dominant source
of decoherence in the considered system can be traced to the presence
of random magnetic fields. Actually, the applicability of this type
of platforms to the realization of fundamental effects in ultracold
atoms depends crucially on shielding the system from magnetic noise
\citep{key-MagneticShield,key-Brouard2FeshbachMagneticNoise}. Quadratic
Zeeman effect, which becomes relevant as the bias field is increased,
was considered in Refs. \citep{key-SpielmanClockTransitions} and
\citep{key-GomezLlorenteNoiseSpectrum}. Since it does not lead to
differential effects in dealing with transverse noise, it is not included
in the present model in order to simplify the analytical treatment.

As appropriate to emulate frequent practical situations, the fluctuations
will be considered to correspond to (Gaussian) Ornstein-Uhlenbeck
processes \citep{key-Gardiner}. Additionally, without loss of generality,
zero mean-values are considered, i.e., 

\begin{eqnarray}
\left\langle \delta\omega_{0}(t)\right\rangle  & = & 0\label{eq:MeanValueInitialDOmeg}\\
\left\langle \eta_{i}(t)\right\rangle  & = & 0,\,i=x,y\label{eq:MeanValuesInitiaEtha}
\end{eqnarray}
 Notice that non-zero mean values can be incorporated as \emph{deterministic}
terms into the model without changing the global structure of the
Hamiltonian. The auto-correlation functions, which have exponential
form as corresponds to OU noise, will be denoted as $G(\delta\omega_{0};\tau)$
and $G(\eta_{i};\tau),\,i=x,y$, i.e., 

\begin{eqnarray}
G(\delta\omega_{0};\tau) & = & \left\langle \delta\omega_{0}(t)\delta\omega_{0}(t+\tau)\right\rangle =\left\langle \delta\omega^{2}_{0}\right\rangle e^{-\alpha_{\delta\omega_{0}}\left|\tau\right|},\\
G(\eta_{i};\tau) & = & \left\langle \eta_{i}(t)\eta_{i}(t+\tau)\right\rangle =\left\langle \eta^{2}_{i}\right\rangle e^{-\alpha_{\eta_{i}}\left|\tau\right|},\,i=x,y,
\end{eqnarray}
 where the parameters $\alpha_{\delta\omega_{0}}$ and $\alpha_{\eta_{i}}$
correspond to the inverses of the respective correlation times, $\tau_{c,\delta\omega_{0}}=1/\alpha_{\delta\omega_{0}}$
and $\tau_{c,\eta_{i}}=1/\alpha_{\eta_{i}}$, of $\delta\omega_{0}(t)$
and $\eta_{i}(t)$. Accordingly, the spectral densities $S(\delta\omega_{0};\omega)$
and $S(\eta_{i};\omega),\,i=x,y$, connected  with the Fourier transform
of the auto-correlation functions through the Wiener-Kinchin theorem
\citep{key-Gardiner,key-Stratonovich}, have Lorentzian form. Namely,
they are given by the expressions

\begin{equation}
S(\delta\omega_{0};\omega)=\frac{\alpha_{\delta\omega_{0}}\left\langle \delta\omega^{2}_{0}\right\rangle }{\pi(\alpha^{2}_{\delta\omega_{0}}+\omega^{2})},
\end{equation}

\begin{equation}
S(\eta_{i};\omega)=\frac{\alpha_{\eta_{i}}\left\langle \eta^{2}_{i}\right\rangle }{\pi(\alpha^{2}_{\eta_{i}}+\omega^{2})},\,i=x,y,\label{eq:SpectrumOU}
\end{equation}
where it is apparent that $\alpha_{\delta\omega_{0}}$ and $\alpha_{\eta_{i}}$
can be considered to give the magnitude of the spectral widths. As
the noise correlation times decrease, the (decaying) spectra become
broader. The white-noise limit, characterized by a Dirac delta auto-correlation
function, and, consequently, by a flat spectrum, will be explicitly
considered. The case of static noise, i.e., of fluctuations with no
time dependence, will also be tackled. Note that, since no restrictions
on the relative magnitude of $\eta_{x}(t)$ and $\eta_{y}(t)$ are
assumed, our model can emulate a realization of anisotropic transverse
noise \citep{key-AnisotropicTransverseNoise}. In particular, the
model can be regarded as corresponding to the incorporation of the
CDD technique into the system realized in Ref. \citep{key-AnisotropicTransverseNoise}.

By now, cross-correlations in the random input are not contemplated,
i.e., it is assumed that 

\begin{eqnarray}
\left\langle \eta_{i}(t)\eta_{j}(t+\tau)\right\rangle  & = & 0,\,i\neq j,\\
\left\langle \eta_{i}(t)\delta\omega_{0}(t+\tau)\right\rangle  & = & 0,\,i=x,y.
\end{eqnarray}
 Further on, some aspects of the potential presence of cross-correlations
in the input and/or of their emergence in the effective stochastic
terms that result from the unitary transformations will be discussed.
Moreover, the generalization of our procedure to incorporate the CDD
concatenation scheme in the description will be outlined. Also, the
possibility of extending our approach to deal with non-Gaussian fluctuations
will be evaluated. Here, it is pertinent to recall that the parameters
found in previous studies as guaranteeing the efficiency of the CDD
technique to cope with longitudinal fluctuations correspond to a high
field intensity, specifically, $\Omega_{d}$ must be much larger than
the noise amplitude, and to a small detuning $\left|\Delta\right|=\left|\omega_{0}-\omega_{d}\right|\ll\omega_{d}$.
One of the objectives of our analysis is to elucidate if, with those
parameters, the control field is useful to curb also the effect of
transverse noise.

\subsection{The noise terms in the dressed-state representation}

Through the unitary transformation

\begin{equation}
\hat{U}_{1}(t)=e^{-i\omega_{d}t\hat{F}_{z}/\hbar},\label{eq:FirstUnitary}
\end{equation}
 and applying the RWA \citep{key-SpielmanClockTransitions}, the Hamiltonian
in Eq. (\ref{eq:BasicHamiltonian}) is rewritten as

\begin{eqnarray}
\hat{H} & = & \left[\Delta+\delta\omega_{0}(t)\right]\hat{F}_{z}+[\Omega_{d}+\chi_{a}(t)]\hat{F}_{x}+\chi_{b}(t)\hat{F}_{y},\label{eq:RotatedBasicHamiltonian}
\end{eqnarray}
 where, for simplicity, the transformed Hamiltonian $\hat{U}^{\dagger}_{1}\hat{H}\hat{U}_{1}-i\hbar\hat{U}^{\dagger}_{1}\dot{\hat{U}}_{1}$
has been denoted $\hat{H}$, as the original one. (Similar compact
notation will be used throughout the paper in different unitary transformations).
Additionally, we have introduced the effective stochastic terms $\chi_{a}(t)$
and $\chi_{b}(t)$, defined by 

\begin{equation}
\chi_{a}(t)=\eta_{x}(t)\cos(\omega_{d}t)+\eta_{y}(t)\sin(\omega_{d}t),\label{eq:DefiningXia}
\end{equation}

\begin{equation}
\chi_{b}(t)=-\eta_{x}(t)\sin(\omega_{d}t)+\eta_{y}(t)\cos(\omega_{d}t).\label{eq:DefiningXib}
\end{equation}
 Now, through the rotation 

\begin{equation}
\hat{U}_{2}(t)=e^{-i\frac{\pi}{2}\hat{F}_{y}/\hbar}\label{eq:Unitary2}
\end{equation}
 the Hamiltonian, in the case of zero detuning, $\Delta=0$, is cast
into the form

\begin{equation}
\hat{H}=[\Omega_{d}+\chi_{a}(t)]\hat{F}_{z}+\delta\omega_{0}(t)\hat{F}_{x}+\chi_{b}(t)\hat{F}_{y},\label{eq:ZeroDetuningHamiltonian}
\end{equation}
 where the term $\Omega_{d}\hat{F}_{z}$ incorporates the transformed
qubit and the rest of elements are random components. The notation
$\left|F,m\right\rangle $, used for the bare states, will be also
employed for their dressed counterparts. The associated (dressed)
eigenenergies are given by $m\hbar\Omega_{d}$. Observe that the fluctuations
$\delta\omega_{0}(t)$, diagonal in the original representation, have
become transverse in the dressed-state basis. Furthermore, the (primarily
orthogonal) random terms $\eta_{x}(t)$ and $\eta_{y}(t)$ have been
cast into a diagonal contribution, $\chi_{a}(t)$, leading to pure
dephasing in the dressed-state basis, and a transverse component,
$\chi_{b}(t)$, which, along with $\delta\omega_{0}(t)$, can induce
population transfers. 

The characteristic properties of $\chi_{a}(t)$ and $\chi_{b}(t)$
are analyzed in the following. It will be shown that the presence
of the (deterministic) oscillating factors in Eqs. (\ref{eq:DefiningXia})
and (\ref{eq:DefiningXib}) can make the spectral densities of $\chi_{a}(t)$
and $\chi_{b}(t)$ to significantly differ from those of $\eta_{x}(t)$
and $\eta_{y}(t)$. Compact results will be achieved via an approximation
consistent with the previously applied RWA.

\subsection{Characterization of the effective noise terms $\chi_{a}(t)$ and
$\chi_{b}(t)$ }

From the definition of $\chi_{a}(t)$ and $\chi_{b}(t)$, and, taking
into account the zero mean values of $\eta_{x}(t)$ and $\eta_{y}(t)$,
one trivially finds 

\begin{equation}
\left\langle \chi_{a}(t)\right\rangle =0
\end{equation}

\begin{equation}
\left\langle \chi_{b}(t)\right\rangle =0
\end{equation}
Additionally, the auto-correlation functions $G(\chi_{a};\tau)$ and
$G(\chi_{b};\tau)$ are obtained as 

\begin{eqnarray}
G(\chi_{a};\tau) & = & \left\langle \chi_{a}(t)\chi_{a}(t+\tau)\right\rangle \nonumber \\
 & = & \frac{1}{2}\left(\left\langle \eta_{x}(t)\eta_{x}(t+\tau)\right\rangle +\left\langle \eta_{y}(t)\eta_{y}(t+\tau)\right\rangle \right)\cos(\omega_{d}\tau)+\nonumber \\
 &  & \frac{1}{2}\left(\left\langle \eta_{x}(t)\eta_{x}(t+\tau)\right\rangle -\left\langle \eta_{y}(t)\eta_{y}(t+\tau)\right\rangle \right)\cos[\omega_{d}(\tau+2t)]\nonumber \\
 & \simeq & \frac{1}{2}\left[G(\eta_{x};\tau)+G(\eta_{y};\tau)\right]\cos(\omega_{d}\tau),\label{eq:CorrelFuncChia}
\end{eqnarray}

\begin{eqnarray}
G(\chi_{b};\tau) & = & \left\langle \chi_{b}(t)\chi_{b}(t+\tau)\right\rangle \nonumber \\
 & = & \frac{1}{2}\left(\left\langle \eta_{x}(t)\eta_{x}(t+\tau)\right\rangle +\left\langle \eta_{y}(t)\eta_{y}(t+\tau)\right\rangle \right)\cos(\omega_{d}\tau)+\nonumber \\
 &  & \frac{1}{2}\left(-\left\langle \eta_{x}(t)\eta_{x}(t+\tau)\right\rangle +\left\langle \eta_{y}(t)\eta_{y}(t+\tau)\right\rangle \right)\cos[\omega_{d}(\tau+2t)]\nonumber \\
 & \simeq & \frac{1}{2}\left[G(\eta_{x};\tau)+G(\eta_{y};\tau)\right]\cos(\omega_{d}\tau),\label{eq:CorrelFuncChib}
\end{eqnarray}
 where, consistently with the previously applied RWA, we have neglected
the terms oscillating with frequency $2\omega_{d}$. Importantly,
this approximation implies regarding $\chi_{a}(t)$ and $\chi_{b}(t)$
as stationary fluctuations \citep{key-Stratonovich}. Note that, in
fact, the terms oscillating with $2\omega_{d}$ strictly vanish for
isotropic transverse noise, i.e., when $G(\eta_{x};\tau)=G(\eta_{y};\tau)$.
Hence, their potential relevance, which goes beyond the RWA, is associated
to the existence of anisotropy in the noisy input. Here, it is worth
noting the analogy with the behavior found in Ref. \citep{key-AnisotropicTransverseNoise}.
There, anisotropy was observed to lead to oscillations out of the
RWA picture. Further on, when dealing with the effect of static noise,
we will show that, with no approximations, an analytical description
of the dynamics can be given beyond the RWA. Then, the parallelism
between the role of anisotropy in the emergence of terms oscillating
with frequency $2\omega_{d}$ in the present setting and the behavior
observed in \citep{key-AnisotropicTransverseNoise} will be soundly
traced. We stress that our description, which does not incorporate
restrictions on the relative magnitude of $\eta_{x}(t)$ and $\eta_{y}(t)$,
is applicable to experimental settings where the fluctuations are
biased. In the case where the noises can be assumed to have equal
magnitude, post-RWA corrections are strictly cancelled.

The effective stochastic terms are correlated, the associated cross-correlation
functions being given by

\begin{equation}
\left\langle \chi_{a}(t)\chi_{b}(t+\tau)\right\rangle =-\frac{1}{2}\left[G(\eta_{x};\tau)+G(\eta_{y};\tau)\right]\sin(\omega_{d}\tau)
\end{equation}

\begin{equation}
\left\langle \chi_{i}(t)\delta\omega_{0}(t+\tau)\right\rangle =0,\,i=a,b
\end{equation}
Moreover, applying the Wiener-Kinchin theorem \citep{key-Gardiner,key-Stratonovich},
the spectral density of $\chi_{a}(t)$ is obtained as 

\begin{eqnarray}
S(\chi_{a};\omega) & = & \frac{1}{2\pi}\int^{\infty}_{-\infty}d\tau e^{-i\omega\tau}G(\chi_{a};\tau)\nonumber \\
 & = & \frac{1}{4\pi}\int^{\infty}_{-\infty}d\tau e^{-i\omega\tau}\left[G(\eta_{x};\tau)+G(\eta_{y};\tau)\right]\cos(\omega_{d}\tau)\nonumber \\
 & = & \frac{1}{4}\left[S(\eta_{x};\omega-\omega_{d})+S(\eta_{x};\omega+\omega_{d})+S(\eta_{y};\omega-\omega_{d})+S(\eta_{y};\omega+\omega_{d})\right].\label{eq:SpectrumXia}
\end{eqnarray}
The same expression is found for the spectral density of $\chi_{b}(t)$,
i.e., 

\begin{equation}
S(\chi_{b};\omega)=S(\chi_{a};\omega).\label{eq:SpectrumXib}
\end{equation}
 Hence, the (correlated) effective random terms that emerge in the
applied rotating system can be considered to have the same characteristics,
provided that the components of their correlation functions oscillating
with frequency $2\omega_{d}$ are neglected. The shifts in $\omega_{d}$
present in the arguments of the spectral densities of $\eta_{x}$
and $\eta_{y}$ as they enter Eq. (\ref{eq:SpectrumXia}) are rooted
in the (deterministic) oscillating factors of Eqs. (\ref{eq:DefiningXia})
and (\ref{eq:DefiningXib}). Those frequency displacements are irrelevant
when $\eta_{x}(t)$ and $\eta_{y}(t)$ have white-noise properties:
the flat character of the spectra implies invariance against shifts
in the frequency. In contrast, in the common case of $\eta_{x}(t)$
and $\eta_{y}(t)$ having decaying spectra, as in the OU processes
considered here, significant variations in the form of $S(\chi_{a};\omega)$
and $S(\chi_{b};\omega)$ can emerge as a consequence of those displacements.
Those features are illustrated in Fig. 1, where it is observed that,
due to the effect of the control field, the dominant part of the spectrum
of the effective noise is displaced to a region around the driving-field
frequency. Since, it is in that spectral range where the fluctuations
can lead to decoherence, the system can be predicted to be noise immune
when its fundamental frequencies are outside that range. The concentration
around $\omega_{d}$ is attenuated as the correlation times $\tau_{c,\eta_{x}}$
and $\tau_{c,\eta_{x}}$ decrease, as can be seen in the right part
of the figure. It will be shown that those qualitative changes in
the spectrum appearance account for a nontrivial role of the effective
fluctuations in the (driven) dynamics.\medskip{}
\medskip{}

\textcolor{red}{\includegraphics[scale=0.6]{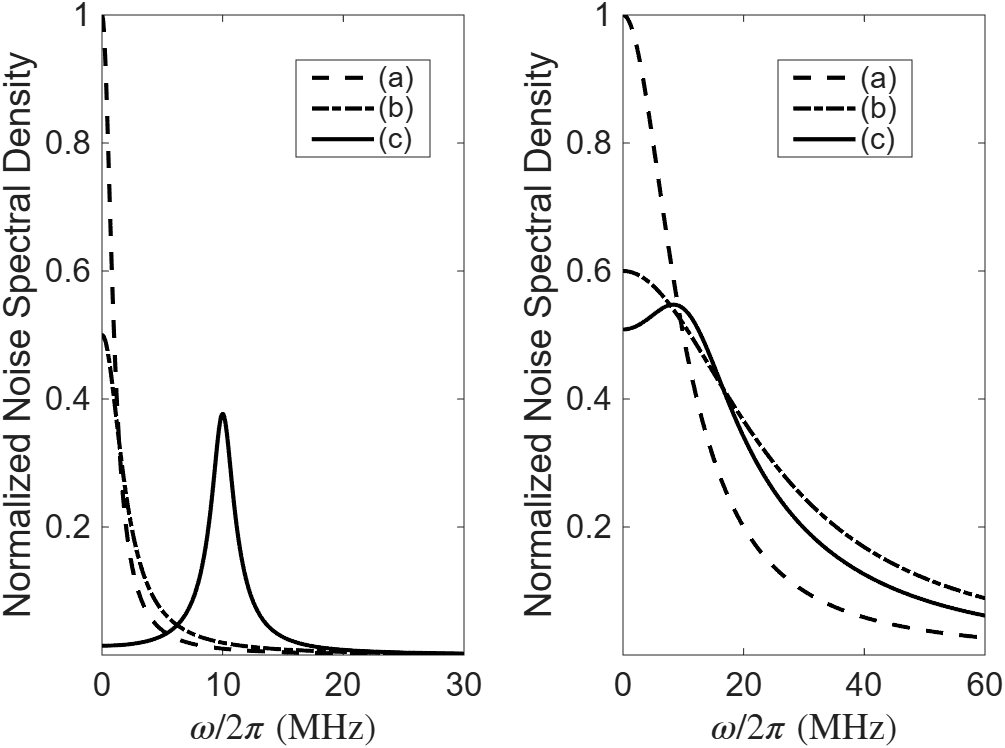}}

\begin{figure}[H]
\caption{Spectral densities of the input fluctuations $\eta_{x}(t)$ (a), and
$\eta_{y}(t)$ (b), and, of the effective noise term $\chi_{a}(t)$
(c). {[}Normalization to the maximum of $S(\eta_{x};\omega)$ has
been used{]}. In the left, the input fluctuations correspond to $\left\langle \eta^{2}_{x}\right\rangle =100\textrm{\,(a.u.)}$,
$\alpha_{\eta_{x}}/2\pi=1\,\textrm{(MHz)}$, $\left\langle \eta^{2}_{y}\right\rangle =100\textrm{(a.u.)}$,
and $\alpha_{\eta_{y}}/2\pi=2\,\textrm{(MHz)}$. In the right, $\left\langle \eta^{2}_{x}\right\rangle =100\textrm{\,(a.u.)}$,
$\alpha_{\eta_{x}}/2\pi=10\,\textrm{(MHz)}$, $\left\langle \eta^{2}_{y}\right\rangle =150\textrm{\,(a.u.)}$,
and $\alpha_{\eta_{y}}/2\pi=25\,\textrm{(MHz)}$. In both, left and
right figures, $\omega_{d}/2\pi=10\,\textrm{(MHz)}$. The variances
are expressed in arbitrary units (a.u.). }

\end{figure}

\subsection*{\textcolor{black}{\bigskip{}
}}

\section{The effect of transverse noise on the efficiency of continuous dynamical
decoupling}

From the analysis of the Hamiltonian in Eq. (\ref{eq:ZeroDetuningHamiltonian}),
it is apparent that, for sufficiently large values of the driving
amplitude $\Omega_{d}$, a perturbative scheme can be applied to study
the effects of noise. Namely, the Hamiltonian can be rewritten as 

\[
\hat{H}=\hat{H}_{0}(t)+\hat{W}(t),
\]
 where the zero-order term is given by
\begin{equation}
\hat{H}_{0}(t)=[\Omega_{d}+\chi_{a}(t)]\hat{F}_{z}.\label{eq:Hcero}
\end{equation}
 and the (time-dependent) perturbation reads 

\begin{eqnarray}
\hat{W}(t) & = & \delta\omega_{0}(t)\hat{F}_{x}+\chi_{b}(t)\hat{F}_{y}.\label{eq:Hperturb}
\end{eqnarray}
 Let us sequentially consider the pure dephasing rooted in the random
component of $\hat{H}_{0}$ and the stochastic transfer of population
between dressed states resulting from the effect of the (transverse)
perturbation.

\subsection{Dressed-state dephasing }

The role of the stochastic component $\chi_{a}(t)\hat{F}_{z}$ in
the dynamics governed by the zero-order Hamiltonian $\hat{H}_{0}$
can be directly characterized applying the methodology presented in
\citep{key-BrouardInternalStateDephasing,key-Paladino1/fNoiseReview}.
In the rotating frame

\begin{equation}
\hat{U}_{3}(t)=e^{-i\Omega_{d}t\hat{F}_{z}/\hbar},
\end{equation}
 the system, prepared in the state $\left|\psi(0)\right\rangle $,
evolves, for each \emph{stochastic trajectory}, as 

\begin{equation}
\left|\psi(t)\right\rangle =e^{-i\zeta_{a}(t)\hat{F}_{z}/\hbar}\left|\psi(0)\right\rangle ,
\end{equation}
where $\zeta_{a}(t)$ is the non-stationary random variable defined
by

\begin{equation}
\zeta_{a}(t)=\int^{t}_{0}\chi_{a}(t^{\prime})dt^{\prime}.\label{eq:FirstNonstatStochVariable}
\end{equation}
 Then, the evolution of a dressed state simply corresponds to a stochastic
variation of the phase. Consequently, the dressed-state populations
do not change. Moreover, the coherences, specifically, the elements
of the density matrix between the states $\left|F,m\right\rangle $
and $\left|F,m^{\prime}\right\rangle $, are shown to evolve as 

\[
\rho_{m,m^{\prime}}(t)=\rho_{m,m^{\prime}}(0)e^{i(m-m^{\prime})\zeta_{a}(t)}.
\]
 Now, incorporating the random character of the system by making the
average over stochastic trajectories, one obtains for the (reduced)
density matrix 

\begin{equation}
\left\langle \rho_{m,m^{\prime}}(t)\right\rangle =\rho_{m,m^{\prime}}(0)\left\langle e^{i(m-m^{\prime})\zeta_{a}(t)}\right\rangle ,\label{eq:expression of the coherences}
\end{equation}
where we have used $\left\langle \right\rangle $ to denote the average
over noise realizations. (Confusion with the standard quantum average,
also denoted as $\left\langle \right\rangle $, will be avoided). 

The applied procedure makes it apparent that the dephasing (in the
dressed-state basis) is not affected by $\delta\omega_{0}(t)$: it
is purely rooted in the fluctuations that enter transversely the original
system. Indeed, this is a consequence of applying the CDD technique:
because of the use of a control field orthogonal\textcolor{red}{{} }to
the qubit, the character of the noise term $\delta\omega_{0}(t)$
changes from diagonal to transverse, and, therefore, no dephasing
of the dressed states results from it. In Appendix C, our description
is generalized by incorporating amplitude noise, i.e., fluctuations
in the driving field amplitude $\Omega_{d}$.

In the following, we will focus on two time regimes, $t\gg\tau_{c}$
and $t\ll\tau_{c}$, ($\tau_{c}$ gives the magnitude of the correlation
times of the involved noises), where a complete analytical characterization
of dephasing is feasible. Actually, the results in those regimes approximate
the generic-time behavior corresponding respectively to broadband
fluctuations, i.e., of noise with $\tau_{c}$ much smaller than the
characteristic time of the system dynamics, and, to the static-noise
scenario, i.e., of fluctuations with large correlation time. 

\subsubsection{The limit of short correlation time}

Following Refs. \citep{key-GomezLlorenteNoiseSpectrum} and \citep{key-Paladino1/fNoiseReview},
it is shown that, for $t\gg\tau_{c}$, the evolution of the coherences
is given by 

\begin{equation}
\left\langle \rho_{m,m^{\prime}}(t)\right\rangle \propto e^{-(m-m^{\prime})^{2}\pi S(\chi_{a};\,\omega=0)t}.\label{PureExponentialDecay}
\end{equation}
 Hence, this regime corresponds to exponential decay, the rate being
determined by the spectral density of $\chi_{a}(t)$ at zero frequency,
$S(\chi_{a};\omega=0)$. Note that these results are particularly
relevant to the white-noise limit ($\tau_{c}\rightarrow0$). 

The dependence of the decay rate on the characteristics of the original
noisy inputs is straightforwardly traced: from Eq. (\ref{eq:SpectrumXia}),
and, taking into account the symmetry of the functions $S(\eta_{x};\omega)$
and $S(\eta_{y};\omega)$, we arrive at 

\begin{equation}
S(\chi_{a};\omega=0)=\frac{1}{2}\left[S(\eta_{x};\omega_{d})+S(\eta_{y};\omega_{d})\right].\label{eq:ExponentPureExponential}
\end{equation}
 At this point, a first conclusion on the effect of transverse noise
on the CDD performance can be drawn. For $\eta_{x}(t)$ and $\eta_{y}(t)$
having declining spectra, $S(\eta_{x};\omega_{d})$ and $S(\eta_{y};\omega_{d})$,
and, in turn, $S(\chi_{a};\omega=0)$, decay as the driving frequency
grows. Therefore, the dephasing is reduced for increasing $\omega_{d}$.
Importantly, we must take into account that an unbound variation of
$\omega_{d}$ as a tool of control is not possible: $\omega_{d}$
must be close to the qubit frequency $\omega_{0}$ to guarantee the
validity of the used approach. Actually, it is the variation of $\omega_{0}$
that might be contemplated in a dephasing-reduction strategy. On the
other hand, for $\eta_{x}(t)$ and $\eta_{y}(t)$ having white-noise
characteristics, and, consequently, flat spectral densities, the dependence
of $S(\chi_{a};\omega=0)$ on $\omega_{d}$ vanishes, and, therefore,
no changes in the dephasing rate result from the modification of the
qubit frequency.

\subsubsection{The static-noise limit}

The term static noise refers to fluctuations with no time dependence:
they take a fixed value in each particular realization, that value
being randomly distributed for the different runs. Combining Eqs.
(\ref{eq:DefiningXia}) and (\ref{eq:FirstNonstatStochVariable}),
the stochastic variable $\zeta_{a}(t)$ can be written in the static
limit as

\begin{eqnarray}
\zeta_{a}(t) & = & \int^{t}_{0}\left[\eta_{x}\cos(\omega_{d}t^{\prime})+\eta_{y}\sin(\omega_{d}t^{\prime})\right]dt^{\prime}\label{eq:FirstStoVariableStaticN}\\
 & = & \eta_{x}\frac{\sin(\omega_{d}t)}{\omega_{d}}-\eta_{y}\frac{\cos(\omega_{d}t)-1}{\omega_{d}}.\nonumber 
\end{eqnarray}
 Consequently, once the average over different random trajectories
is carried out {[}see Eq. (\ref{eq:expression of the coherences}){]},
we find, after some algebra, that the coherences are given by

\begin{eqnarray}
\left\langle \rho_{m,m^{\prime}}(t)\right\rangle  & = & \rho_{m,m^{\prime}}(0)e^{-\frac{(m-m^{\prime})^{2}}{\omega^{2}_{d}}\left(\frac{\left\langle \eta^{2}_{x}\right\rangle +\left\langle \eta^{2}_{y}\right\rangle }{2}\left[1-\cos(\omega_{d}t)\right]+\frac{\left\langle \eta^{2}_{x}\right\rangle -\left\langle \eta^{2}_{y}\right\rangle }{4}\left[2\cos(\omega_{d}t)-\cos(2\omega_{d}t)-1\right]\right)}.\label{eq:StaticDephasing}
\end{eqnarray}
 Since this analysis transcends the RWA, (we have retained terms oscillating
with frequency $2\omega_{d}$), its consistency demands the generalization
of the description: to assess the actual relevance of the found (noise-induced)
evolution of the coherences, we must incorporate into the description
the corrections to the RWA coming from the \emph{deterministic} part
of the Hamiltonian. The study of those effects, presented in Appendix
A, allows assessing the relative importance of the two contributions
(deterministic and stochastic) to the post RWA dynamics. From the
inspection of Eq. (\ref{eq:StaticDephasing}), it follows that the
magnitude of the noisy oscillations in the exponent is approximately
given by $\left\langle \eta^{2}_{i}\right\rangle /\omega^{2}_{d}$;
in contrast, as can be seen in Appendix A, the order of the deterministic
analogues corresponds to $\Omega_{d}/\omega_{d}$. Then, assuming
that the longitudinal and transverse fluctuations have similar magnitude,
i.e., $\left\langle \delta\omega^{2}_{0}\right\rangle \sim\left\langle \eta^{2}_{i}\right\rangle $,
it is concluded that, in the standard CDD scenario, where the restrictions
$\Omega_{d}\ll\omega_{d}$, and, $\delta\omega_{0}\ll\Omega_{d}$
must be fulfilled, the stochastic contribution to the coherence evolution
is in fact secondary to its deterministic counterpart. No significant
dephasing results from transverse static noise in that regime of parameters.
Notice that the applied framework allows us to analyze how these conclusions
are modified as variations of the relative magnitude of the longitudinal
and transverse fluctuations are considered.

Here, it is worth going beyond the standard CDD scenario and discussing
the predictions of Eq. (\ref{eq:StaticDephasing}) for systems where
higher values of the quotient $\left\langle \eta^{2}_{i}\right\rangle /\omega^{2}_{d}$
are reached. From the oscillating character of the exponent, one can
soundly conclude that, for a generic value of the oscillation amplitude
$z\sim\left\langle \eta^{2}_{i}\right\rangle /\omega^{2}_{d}$,  a
complex evolution of the coherences can occur: a variety of terms
oscillating with frequencies multiples of those contained in the exponent
contribute to the $z$-power expansion of $\left\langle \rho_{m,m^{\prime}}(t)\right\rangle $.
Note that oscillations with multiple frequencies emerge irrespective
of the isotropic or anisotropic character of the fluctuations. As
the magnitude of $z$ decreases, considerable simplification comes
about: for $z\ll1$, the dominant oscillating contribution to the
coherences can be traced to the (fundamental) frequencies present
in the exponent. In this regime, the differential effects of anisotropic
noise become evident: whereas, in the case of isotropic fluctuations
($\left\langle \eta^{2}_{x}\right\rangle =\left\langle \eta^{2}_{y}\right\rangle $),
the oscillations of the exponent are characterized by a single frequency
$\omega_{d}$, for anisotropic noise, a contribution with frequency
$2\omega_{d}$ is also present. These findings are illustrated in
Fig. 2. Note that the chosen parameters are outside the range of applicability
of the CDD. As expected, for isotropic noise, the coherences are observed
to evolve with a single frequency ($\omega_{d}$). In contrast, in
the case of anisotropic fluctuations, the evolution incorporates components
with frequencies $\omega_{d}$ and $2\omega_{d}$. These results parallel
the \emph{frequency doubling} detected in the scenario implemented
in Ref. \citep{key-AnisotropicTransverseNoise}, where different forms
of anisotropic fluctuations were synthetically generated. There, no
driving field was incorporated, and, it was the qubit frequency that
was observed to duplicate in the system response to anisotropic static
noise. It is worth stressing that, although our results are strictly
valid only in the static-noise limit, in our approach, we had previously
traced the differential role of anisotropy in a general-noise regime.
Specifically, anisotropy was associated to the emergence of terms
oscillating with frequency $2\omega_{d}$ in the properties of the
effective fluctuations {[}see Eq. (\ref{eq:CorrelFuncChia}){]}. \medskip{}
\medskip{}

\includegraphics[scale=0.6]{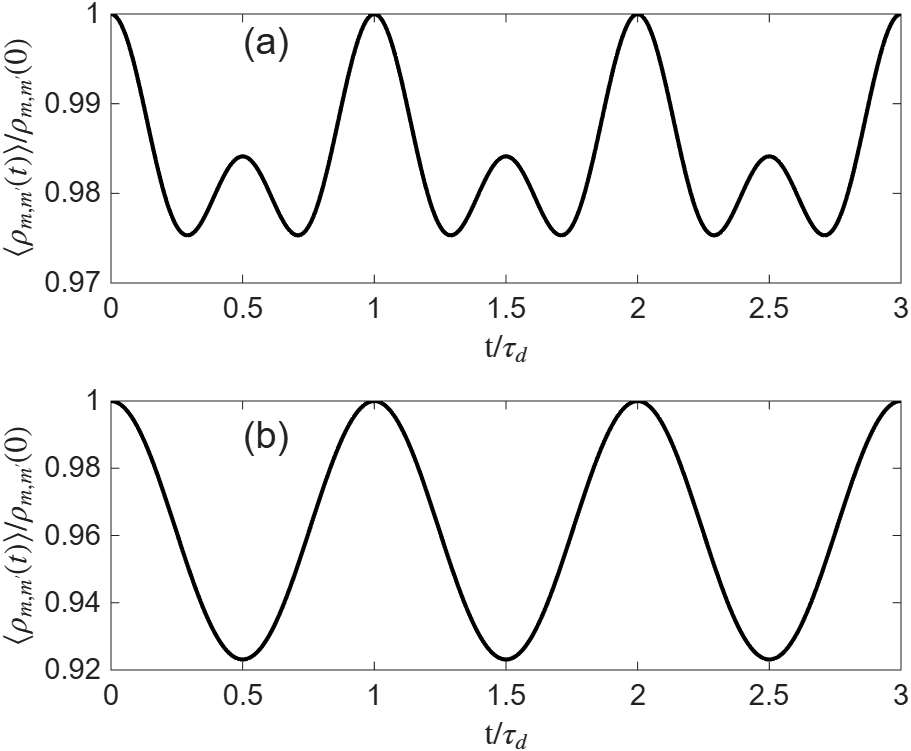}

\begin{figure}[H]

\caption{Dephasing resulting from anisotropic (a) and isotropic (b) static
noisy inputs. (a) $\left\langle \eta^{2}_{x}\right\rangle =1$, $\left\langle \eta^{2}_{y}\right\rangle =0.2$,
$\omega_{d}=5$. (b) $\left\langle \eta^{2}_{x}\right\rangle =\left\langle \eta^{2}_{y}\right\rangle =1$,
$\omega_{d}=5$. Arbitrary units are used for the variances and the
frequency. We have taken $\left|m-m^{\prime}\right|=1$.}

\end{figure}

\subsection{Transfer of population between the dressed states }

Now, we turn to incorporate the perturbation into the description.
Applying time-dependent perturbation theory, it is shown that, for
a particular stochastic trajectory, the probability $P_{m,m^{\prime}}(t)$
for a noise-induced transition between two dressed states $\left|F,m\right\rangle $
and $\left|F,m^{\prime}\right\rangle $ is given by 

\begin{eqnarray*}
P_{m,m^{\prime}}(t) & = & \frac{1}{\hbar^{2}}\left|\int^{t}_{0}dt^{\prime}W_{m,m^{\prime}}(t^{\prime})e^{i(m-m^{\prime})[\Omega_{d}t^{\prime}+\zeta_{a}(t^{\prime})]}\right|^{2}.
\end{eqnarray*}
 Moreover, by averaging over stochastic realizations, and, taking
into account that no cross-correlations between $\delta\omega_{0}(t)$
and $\chi_{b}(t)$ are assumed, we arrive at

\begin{eqnarray}
\left\langle P_{m,m^{\prime}}(t)\right\rangle  & = & \mathcal{F}^{(x)}_{m,m^{\prime}}\left\langle \left|\int^{t}_{0}dt^{\prime}\delta\omega_{0}(t^{\prime})e^{i(m-m^{\prime})[\Omega_{d}t^{\prime}+\zeta_{a}(t^{\prime})]}\right|^{2}\right\rangle +\nonumber \\
 &  & \mathcal{F}^{(y)}_{m,m^{\prime}}\left\langle \left|\int^{t}_{0}dt^{\prime}\chi_{b}(t^{\prime})e^{i(m-m^{\prime})[\Omega_{d}t^{\prime}+\zeta_{a}(t^{\prime})]}\right|^{2}\right\rangle \label{eq:GeneralAverPopTransfer}
\end{eqnarray}
where we have introduced the state-dependent coefficients

\begin{eqnarray}
\mathcal{F}^{(x)}_{m,m^{\prime}} & = & \frac{\left|\left\langle F,m\left|\hat{F}_{x}\right|F,m^{\prime}\right\rangle \right|^{2}}{\hbar^{2}},\nonumber \\
\mathcal{F}^{(y)}_{m,m^{\prime}} & = & \frac{\left|\left\langle F,m\left|\hat{F}_{y}\right|F,m^{\prime}\right\rangle \right|^{2}}{\hbar^{2}},
\end{eqnarray}
 which are determined by the system characteristics.

As in the analysis of the dephasing, it is convenient to differentiate
here the results corresponding to broadband noise and to static fluctuations.
In the following, those cases are separately tackled.

\subsubsection{The limit of short correlation time}

Following the procedure introduced in Ref. \citep{key-GomezLlorenteNoiseSpectrum},
one obtains that, for $t\gg\tau_{c}$, the transfer probability averaged
over noise realizations can be approximated as (see Appendix B) 

\begin{eqnarray}
\left\langle P_{m,m^{\prime}}(t)\right\rangle  & = & 2\pi t\left[\mathcal{F}^{(x)}_{m,m^{\prime}}S(\delta\omega_{0};\tilde{\Omega}_{d})+\mathcal{F}^{(y)}_{m,m^{\prime}}S(\chi_{b};\tilde{\Omega}_{d})\right].\label{eq:AveragedPopulTransfer}
\end{eqnarray}
 where $\tilde{\Omega}_{d}$ denotes the effective frequency corresponding
to the considered transition, i.e., 
\begin{equation}
\tilde{\Omega}_{d}=\left|(m-m^{\prime})\Omega_{d}\right|,
\end{equation}
which, importantly, is proportional to the driving-field amplitude
$\Omega_{d}$. In the derivation of Eq. (\ref{eq:AveragedPopulTransfer}),
we have taken into account that $\Omega_{d}t\gg\zeta_{a}(t)$, and,
as a consequence, we have neglected the random shift in the frequency,
associated to the previously characterized dephasing. That stochastic
displacement has a second-order effect on the transfer of population. 

The key role played by the spectral densities of $\delta\omega_{0}(t)$
and $\chi_{b}(t)$ in the transition is apparent in Eq. (\ref{eq:AveragedPopulTransfer}):
the magnitude of the transfer rate is determined by the values of
the spectral densities at the effective frequency $\tilde{\Omega}_{d}$.
A detailed analysis of the found expression uncovers more specific
features: 

i) In the case of fluctuations with decaying spectra, the values of
the driving amplitude and frequency can strongly affect the magnitude
of the transfer rate. Let us first consider the role of the field
amplitude. From the decay of $S(\delta\omega_{0};\tilde{\Omega}_{d})$
with $\Omega_{d}$, one concludes that the probability for $\delta\omega_{0}(t)$
to induce a transition can be significantly reduced by working with
sufficiently large values of $\Omega_{d}$. Similar implications has
the decay of $S(\chi_{b};\tilde{\Omega}_{d})$ with $\Omega_{d}$.
Obviously, the restrictions existent on the field intensity for the
used description to be valid must be taken into account: the applicability
of the RWA imposes the condition $\omega_{d}\gg\Omega_{d}$. Indeed,
this restriction is decisive in characterizing the role of $\omega_{d}$:
given that $S(\chi_{b};\tilde{\Omega}_{d})$ is determined by $S(\eta_{x};\tilde{\Omega}_{d}\pm\omega_{d})$
and $S(\eta_{y};\tilde{\Omega}_{d}\pm\omega_{d})$, it can be significantly
turned down as the magnitude of $\omega_{d}$ is increased. Specifically,
 for $\omega_{d}\gg\Omega_{d}$, negligible values of $S(\chi_{b};\tilde{\Omega}_{d})$
are reached for noise with sufficiently large correlation times. (Here,
it is worth stressing that we are dealing with the time regime $t\gg\tau_{c}$.
Hence, our conclusions apply to a scenario where the fluctuations
are outside the white-noise range, i.e., $\tau_{c}$ is far from the
limit $\tau_{c}\rightarrow0$ , and still it is much smaller that
the time $t$ considered in the system evolution). From the associated
reduction in the transfer rate, it follows that no significant transfer
of population results from transverse noise in the standard range
of parameters of the CDD technique. Here, it is pertinent to recall
that to guarantee the validity of the used approach, $\omega_{d}$
must be close to the qubit frequency $\omega_{0}$. Consequently,
an unbound variation of $\omega_{d}$ is not feasible as a tool of
control.

ii) As the white-noise limit is approached, the above arguments lose
applicability. Since the (flat) spectral densities are not modified
as $\tilde{\Omega}_{d}$ is varied, the changes in the amplitude of
the CDD field of control have no use in curbing the noise-induced
transitions. No effect on decoherence reduction results either from
varying $\omega_{d}$. 

Valuable insight into general mechanisms responsible for the found
behavior is obtained using a harmonic expansion of the fluctuations,
(see for instance Ref. \citep{key-Gardiner}). In that picture, the
noisy Hamiltonian can be regarded as corresponding to a multiplet
non-diagonally driven by a superposition of harmonic signals with
a continuous set of frequencies, the weights of the different (random)
components being distributed according to the noise spectral density.
Well-known results of the study of harmonically-driven systems can
be used to interpret general features. One can understand that only
the signals with frequencies close to the system characteristic frequencies
are able to induce significant changes of population. Moreover, the
associated probabilities are known to be proportional to the spectral
density at the involved quasiresonant frequencies. In the case of
white noise, the different components have the same weight in the
expansion. Then, no changes occur as the transition frequency is modified.
In contrast, for decaying spectra, as the transition frequency grows,
the impact of the quasiresonant harmonic components becomes negligible.
From this picture, the CDD techniques can be regarded as based on
taking the effective transition frequency out of the noise spectral
range. It is then understood that the static-noise regime, characterized
by narrow spectra centered on zero frequency, is a particularly appropriate
setting for the applicability of the method.

\subsubsection{The static-noise limit}

In the case of fluctuations with no time dependence, Eq. (\ref{eq:GeneralAverPopTransfer})
is developed to obtain 

\begin{eqnarray}
\left\langle P_{m,m^{\prime}}(t)\right\rangle  & = & \mathcal{F}^{(x)}_{m,m^{\prime}}\left\langle \delta\omega^{2}_{0}\right\rangle \frac{\sin^{2}(\frac{\tilde{\Omega}_{d}t}{2})}{(\tilde{\Omega}_{d}/2)^{2}}+\nonumber \\
 &  & \mathcal{F}^{(y)}_{m,m^{\prime}}\left(\left\langle \eta^{2}_{x}\right\rangle +\left\langle \eta^{2}_{y}\right\rangle \right)\left[\frac{\sin^{2}(\frac{\Omega_{+}t}{2})}{\Omega^{2}_{+}}+\frac{\sin^{2}(\frac{\Omega_{-}t}{2})}{\Omega^{2}_{-}}\right]+\label{eq:StaticLimitTransfer}\\
 &  & \mathcal{F}^{(y)}_{m,m^{\prime}}\left(-\left\langle \eta^{2}_{x}\right\rangle +\left\langle \eta^{2}_{y}\right\rangle \right)\left[\frac{\cos^{2}(\omega_{d}t)+\cos(\tilde{\Omega}_{d}t)\cos(\omega_{d}t)}{\Omega_{+}\Omega_{-}}\right]\nonumber 
\end{eqnarray}
 where $\Omega_{+}=\tilde{\Omega}_{d}+\omega_{d}$, and $\Omega_{-}=\tilde{\Omega}_{d}-\omega_{d}$.
Similarly to the behavior observed in the analysis of dephasing, we
find here that the existence of anisotropy in the original noisy inputs
affects the oscillatory character of the transition probability. Namely,
for anisotropic quasistatic noise, the Fourier components of the found
oscillatory behavior includes a contribution at the frequency $2\omega_{d},$
apart from terms at frequencies $\tilde{\Omega}_{d}$ and $\omega_{d}$,
also present in the isotropic case. Considering the different noisy
inputs to have similar magnitudes, i.e., $\left\langle \delta\omega^{2}_{0}\right\rangle \sim\left\langle \eta^{2}_{i}\right\rangle $,
it is concluded that, in the standard range of parameters where the
CDD is applicable, that contribution, which has amplitude approximately
given by $\left\langle \eta^{2}_{i}\right\rangle /\omega^{2}_{d}$,
has second-order character compared with its deterministic counterpart,
of amplitude $\mathcal{F}^{(y)}_{m,m^{\prime}}\Omega^{2}_{d}/\omega^{2}_{d}\sim\Omega^{2}_{d}/\omega^{2}_{d}$,
(see Appendix A). The implications of varying the relative magnitude
of the noises can be traced in our framework.

We have left out of our general description some aspects of the system
which were considered to be not relevant to the differential role
of transverse noise in the CDD efficiency. Some reviewing comments
on them are pertinent:

i) A qualitative estimation of the role of the potential existence
of correlations between the considered sources of noise can be given.
The existence of correlations of $\delta\omega_{0}(t)$ with $\eta_{x}(t)$
and/or $\eta_{y}(t)$, which, in turn, leads to cross-correlations
between $\delta\omega_{0}(t)$ and $\chi_{b}(t)$, implies an additional
contribution to $\left\langle P_{m,m^{\prime}}(t)\right\rangle $.
Again, the presence of oscillating factors in the cross-correlation
function gives relevance to the previous arguments on the reduced
magnitude of that contribution to the random induced transitions.
Hence, no significant modification of the CDD performance can be expected
from the existence of cross-correlations.

ii) One can additionally predict that the concatenation scheme employed
to cope with the extra noise introduced by the field of control does
not lead to differential elements in the analysis of transverse noise.
Indeed, in the application of the CDD method one can always consider
the following procedure. In a first step, the effect of the primary
sources of noise (longitudinal and transverse) can be curbed via a
convenient choice of the system parameters. Subsequently, the technique
can proceed by dealing with the extra fluctuations irrespective of
the character of the noise addressed in the former step. Actually,
the presence of transverse fluctuations does not affect the applicability
of the protocols designed to deal with extra noise. Here, one must
note that amplitude fluctuations can be incorporated into our basic
framework as an additional contribution to transverse noise. The procedure
is quite simple if the fluctuations can be assumed to have static
character. In Appendix C, we extend our description along that line.

iii) It is also worth pondering the relevance that the Gaussian characteristics
assumed for the fluctuations have in our description. Since the application
of time-dependent perturbation theory to first-order requires only
up to the second moment of noise, the used framework embodies in fact
a Gaussian approximation. Hence, the analysis of the role of non-Gaussian
fluctuations demands the generalization of the approach \citep{key-Viola4NonGaussian}.

\section{Numerical study}

We present in this section the results of a numerical simulation of
the dephasing process described by Eq. (\ref{eq:expression of the coherences}).
Our objective is to give numerical support to our analytical results.
We will show that, as predicted by the theory, the efficiency of the
CDD method to curb decoherence induced by transverse noise grows as
the driving frequency is increased and as the relevant fluctuations
have smaller spectral widths. More specifically, we will confirm the
accuracy of the analytical expressions previously derived to describe
the dephasing process.

The characteristics of the fluctuations present in the different settings
proposed for quantum information are widely varied. Consequently,
there is a broad variety of noise parameters that can be employed
to emulate practical situations. Here, we have chosen a set of parameters
that might be relevant to the scenario studied in \citep{key-SpielmanClockTransitions}.
As in our analytical approach, we have considered uncorrelated noisy
inputs $\eta_{x}(t)$ and $\eta_{y}(t)$ with Ornstein-Uhlenbeck characteristics
and different correlation times. Specifically, for the noise standard
deviation, we have taken $\sqrt{\left\langle \eta^{2}_{i}\right\rangle }/2\pi\sim0.1\textrm{\,(MHz)}$
($i=x,y$). Moreover, the noise spectral width $\alpha_{\eta_{i}}$
is assumed to be in the range ($1-10)\textrm{\,(MHz)}$. Additionally,
different values of the driving frequency $\omega_{d}$ in the range
$\omega_{d}/2\pi\sim10\,\textrm{(MHz)}$ have been used. We have applied
standard techniques in the numerical simulation of the stochastic
processes \citep{key-Gillespie}. Our results are presented in Figure
3. The black lines represent the results of the numerical study; the
red lines correspond to the theoretical predictions. The left part
of the figure corresponds to fluctuations with larger correlation
times, i.e., smaller spectral width, than their counterparts in the
right. To isolate the differential role of the spectral width, we
have used noises with the same variances in both (left and right)
figures. From the results, it is apparent that the CDD efficiency
increases as the spectral widths of the fluctuations are reduced:
the decoherence processes depicted in the left are much slower than
those in the right. (Observe the different vertical ranges covered
in the two figures). It is also evident from both figures that decoherence
is increasingly curbed as larger driving frequencies are employed.
{[}Notice that growing values of $\omega_{d}$ are considered from
(a) to (c){]}. It is important to take into account that, given the
magnitude of the involved correlation times $\tau_{c}$, the vast
majority of the represented time range corresponds to the regime $t\gg\tau_{c}$.
Hence, consistently with the prediction of Eq. (\ref{PureExponentialDecay}),
a simple exponential decay is observed. Furthermore, it is shown that
the exponent obtained via the numerical calculations accurately fits
the expression $\pi S(\chi_{a};\,\omega=0)$, given by Eq. (\ref{eq:ExponentPureExponential}).
Actually, combining Eqs. (\ref{eq:SpectrumOU}) and (\ref{eq:ExponentPureExponential}),
it follows that the dephasing time $T_{2}$ is given by the expression
\[
T_{2}=\left[\pi S(\chi_{a};\,\omega=0)\right]^{-1}=\left(\frac{1}{2}\left[\frac{\alpha_{\eta_{x}}\left\langle \eta^{2}_{x}\right\rangle }{(\alpha^{2}_{\eta_{x}}+\omega^{2}_{d})}+\frac{\alpha_{\eta_{y}}\left\langle \eta^{2}_{y}\right\rangle }{(\alpha^{2}_{\eta_{y}}+\omega^{2}_{d})}\right]\right)^{-1},
\]
 which agrees with the numerical results corresponding to all the
situations considered in Fig. 3. Importantly, these results make it
apparent how the dependence of the decoherence process on the noise
characteristics and on the driving frequency can be precisely traced
in our framework. Convenient for a simple illustration of the CDD
efficiency is to give some numerical values. For the cases corresponding
to Fig. 3 {[}left, (b){]} and Fig. 3 {[}right, (b){]}, we respectively
obtain $T_{2}\sim6\times10^{3}\mu s$ and $T_{2}\sim6\times10^{2}\mu s$.

The found agreement between theory and numerical simulation, achieved
without incorporating adjustable parameters in our analytical description,
makes it apparent that the dephasing process is accurately described
in our study. 

\medskip{}
\medskip{}

\includegraphics[clip,scale=0.6]{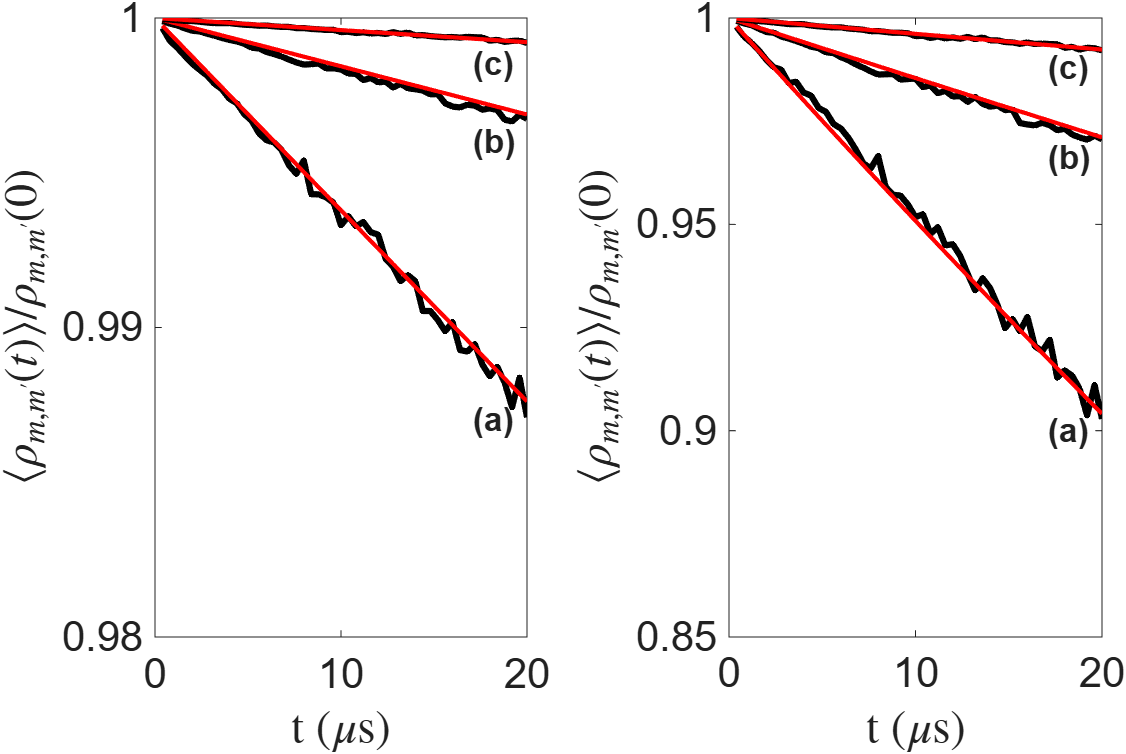}

\begin{figure}[H]
\caption{Results of the numerical simulation of the dephasing process (black
lines) as compared with the predictions of the theoretical approach
(red lines). In the left part, $\alpha_{\eta_{x}}=1\,\textrm{(MHz)}$,
$\alpha_{\eta_{y}}=0.5\,\textrm{(MHz)}$; in the right part, $\alpha_{\eta_{x}}=10\,\textrm{(MHz)}$,
$\alpha_{\eta_{y}}=5\,\textrm{(MHz)}$. In both (left and right) parts,
$\sqrt{\left\langle \eta^{2}_{x}\right\rangle }/2\pi=0.1\textrm{\,(MHz)}$,
$\sqrt{\left\langle \eta^{2}_{y}\right\rangle }/2\pi=0.15\textrm{(MHz)}$,
and, $\omega_{d}/2\pi=5\,\textrm{(MHz)}$ (a); $\omega_{d}/2\pi=10\,\textrm{(MHz)}$
(b); and $\omega_{d}/2\pi=20\,\textrm{(MHz)}$ (c).}
\end{figure}

\section{Adiabatic preparation of the dressed states: The relevance of noisy
Landau-Zener transitions}

\subsection{Obtaining the dressed states}

The dressed states are frequently prepared applying adiabatic passage
techniques. Specifically, with the system in an eigenstate of the
undriven Hamiltonian, the control field is adiabatically switched
on: the amplitude, incorporated into $\Omega_{d}$, and the detuning
$\Delta$ are conveniently varied through linear ramps at sufficiently
slow rates till reaching the values appropriate for the CDD technique
to be effective. Namely, the ramps come to an end around $\Delta=0$,
and, when a sufficiently large value of $\Omega_{d}$ has been attained.
In this form, the system is made to follow the adiabatic eigenstate
that initially corresponds to the prepared bare state and that ends
up as the (dressed) target. The objective of this section is to evaluate
the robustness of this method against the presence of fluctuations
in the setup. Note that, in the platform implemented in Ref. \citep{key-SpielmanClockTransitions},
the detuning, $\Delta=\omega_{0}-\omega_{d}$, was modified by changing
the bias magnetic field that determines the Zeeman multiplet frequency
$\omega_{0}$.

We consider that the dynamics of the system is governed by the Hamiltonian
in Eq. (\ref{eq:RotatedBasicHamiltonian}), where a linear variation
of the control-field parameters is realized. In the experimental arrangements,
(see for instance Ref. \citep{key-SpielmanClockTransitions} and references
therein), combined changes of $\Delta$ and $\Omega_{d}$ are implemented:
the final values of those parameters are reached via ramps formed
by a sequence of different steps; at each step, one of the parameters
is kept constant whereas the other is varied. Here, without loss of
generality, the simultaneous modification of both parameters will
be assumed. We will proceed by obtaining first the adiabatic states
of the Hamiltonian in Eq. (\ref{eq:RotatedBasicHamiltonian}). Subsequently,
we will evaluate the effect of the fluctuations on them. The description
is considerably simplified through the unitary transformation

\begin{equation}
\hat{U}_{4}(t)=e^{-i\theta(t)\hat{F}_{y}/\hbar},
\end{equation}
where $\theta$ is defined by 

\begin{equation}
\theta(t)=\arctan\left[\frac{\Omega_{d}(t)}{\Delta(t)}\right].
\end{equation}
 The transformed Hamiltonian reads 

\begin{eqnarray}
\hat{H} & = & \left[\sqrt{\Delta(t)^{2}+\Omega^{2}_{d}(t)}+\chi_{c}(t)\right]\hat{F}_{z}+\chi_{d}(t)\hat{F}_{x}+\chi_{b}(t)\hat{F}_{y},\label{eq:preparationHamiltonian}
\end{eqnarray}
 where, in order to provide a compact characterization of the (transformed)
stochastic components, we have introduced the effective random terms
\begin{eqnarray}
\chi_{c}(t) & = & \delta\omega_{0}(t)\cos\theta+\chi_{a}(t)\sin\theta,\label{eq:DefXic}\\
\chi_{d}(t) & = & -\delta\omega_{0}(t)\sin\theta+\chi_{a}(t)\cos\theta.\label{eq:DefXid}
\end{eqnarray}
Due to the time dependence of $\Delta$ and $\Omega_{d}$, there are
additional contributions to the transformed Hamiltonian. Those extra
terms, which are proportional to the rate of change of the parameters,
can be neglected in the considered slow-variation regime. 

From the analysis of the \emph{deterministic} part of the Hamiltonian
in Eq. (\ref{eq:preparationHamiltonian}), it is apparent that the
adiabatic states are the eigenstates of $F_{z}$. Their time dependence
becomes explicit by going back in the sequence of unitary transformations.
The associated (adiabatic) eigenvalues $E^{(AD)}_{m}(t)$ are given
by 

\begin{equation}
E^{(AD)}_{m}(t)=m\hbar\sqrt{\Delta(t)^{2}+\Omega^{2}_{d}(t)}
\end{equation}
As required for the evaluation of the potential occurrence of noise-induced
transitions between adiabatic states, we present in the following
the analysis of the properties of $\chi_{c}(t)$ and $\chi_{d}(t)$. 

\subsection{Characterization of the effective noise terms $\chi_{c}(t)$ and
$\chi_{d}(t)$}

From the definition of $\chi_{c}(t)$ and $\chi_{d}(t)$, their mean
values are directly obtained as 

\subsection*{
\begin{eqnarray}
\left\langle \chi_{c}(t)\right\rangle  & = & 0,\\
\left\langle \chi_{d}(t)\right\rangle  & = & 0.
\end{eqnarray}
}

Moreover, for the correlation functions one finds 
\begin{eqnarray}
G(\chi_{c};\tau) & = & \left\langle \chi_{c}(t)\chi_{c}(t+\tau)\right\rangle \nonumber \\
 & = & \left\langle \delta\omega_{0}(t)\delta\omega_{0}(t+\tau)\right\rangle \cos^{2}\theta+\left\langle \chi_{a}(t)\chi_{a}(t+\tau)\right\rangle \sin^{2}\theta\nonumber \\
 & = & G(\delta\omega_{0};\tau)\cos^{2}\theta+G(\chi_{a};\tau)\sin^{2}\theta,
\end{eqnarray}

\begin{eqnarray}
G(\chi_{d};\tau) & = & \left\langle \chi_{d}(t)\chi_{d}(t+\tau)\right\rangle \nonumber \\
 & = & \left\langle \delta\omega_{0}(t)\delta\omega_{0}(t+\tau)\right\rangle \sin^{2}\theta+\left\langle \chi_{a}(t)\chi_{a}(t+\tau)\right\rangle \cos^{2}\theta\nonumber \\
 & = & G(\delta\omega_{0};\tau)\sin^{2}\theta+G(\chi_{a};\tau)\cos^{2}\theta,
\end{eqnarray}
 and, applying the Wiener-Kinchin theorem, the associated spectral
densities are shown to be given by 
\begin{equation}
S(\chi_{c};\omega)=S(\delta\omega_{0};\omega)\cos^{2}\theta+S(\chi_{a};\omega)\sin^{2}\theta,\label{eq:SpectrumXic}
\end{equation}

\begin{equation}
S(\chi_{d};\omega)=S(\delta\omega_{0};\omega)\sin^{2}\theta+S(\chi_{a};\omega)\cos^{2}\theta.\label{eq:SpectrumXid}
\end{equation}
 It is assumed that the slow time variation of $\theta$ does not
affect the \emph{stationary} character of the effective fluctuations.
Note that $\omega_{d}$ enters indirectly Eqs. (\ref{eq:SpectrumXic})
and (\ref{eq:SpectrumXid}) through $S(\chi_{a};\omega)$. Hence,
all the considerations previously made on the shift in the argument
of the spectral density induced by the driving field are still applicable.

The combination of noises present in the definition of $\chi_{c}(t)$
and $\chi_{d}(t)$ leads also to the appearance of cross-correlation,
namely, 

\begin{eqnarray}
\left\langle \chi_{c}(t)\chi_{d}(t+\tau)\right\rangle  & = & \left[-G(\delta\omega_{0};\tau)+G(\chi_{a};\tau)\right]\cos\theta\sin\theta,\\
\left\langle \chi_{c}(t)\chi_{b}(t+\tau)\right\rangle  & = & -\frac{1}{2}\left[G(\eta_{x};\tau)+G(\eta_{y};\tau)\right]\sin(\omega_{d}\tau)\sin\theta+\nonumber \\
 &  & \frac{1}{2}\left[-G(\eta_{x};\tau)+G(\eta_{y};\tau)\right]\sin[\omega_{d}(2t+\tau)]\sin\theta\\
\left\langle \chi_{d}(t)\chi_{b}(t+\tau)\right\rangle  & = & -\frac{1}{2}\left[G(\eta_{x};\tau)+G(\eta_{y};\tau)\right]\sin(\omega_{d}\tau)\cos\theta+\nonumber \\
 &  & \frac{1}{2}\left[-G(\eta_{x};\tau)+G(\eta_{y};\tau)\right]\sin[\omega_{d}(2t+\tau)]\cos\theta
\end{eqnarray}

Important for the analysis of the dynamical effect of the cross-correlation
is the presence of the deterministic oscillating factor in the above
expressions. It is also apparent that terms oscillating with frequency
$2\omega_{d}$ vanish in the case of isotropic noisy inputs.

\subsection{Preparing the dressed states}

To illustrate the basis of the preparation method, we represent in
Fig. 4 the adiabatic (dressed) eigenvalues $E^{(AD)}_{m}(t)$ and
their diabatic (bare) counterparts $E^{(D)}_{m}(t)=m\hbar\Delta(t)$
corresponding to a step in the procedure where $\Omega_{d}$ is fixed
and $\Delta$ is being linearly varied. Furthermore, to simplify the
identification of the mechanism responsible for the effectiveness
of the scheme, we present a compact picture of the process: without
loss of generality, we consider that, in Fig. 4, $\Omega_{d}$ has
already reached its final value, (i.e., the value required for the
CDD to be operative). Notice that the bare and the dressed energy
levels characterized by $m=0$ are equal (horizontal line). Observe
also that $\Omega_{d}$ corresponds to half the separation between
the highest and the lowest adiabatic levels at $\Delta=0$, i.e.,
at the (avoided) crossing. One can see that for $\left|\Delta\right|\gg\Omega_{d}$,
the dressed states approach the eigenstates of the undriven Hamiltonian.
Moreover, the system, which, in the absence of noise, is described
by the Hamiltonian $\Delta\hat{F}_{z}+\Omega_{d}\hat{F}_{x}$, incorporates
the basic ingredients of the Landau-Zener (LZ) model \citep{key-Zener,key-Landau},
\citep{key-LZ3Zenesini,key-CarrollHioeLZ3Analytical,key-BandAvishaiLZ3}.
Namely, the diabatic energy-level differences are linearly varied,
and, additionally, at the starting point $\left|\Delta\right|\gg\Omega_{d}$,
the system is far from the crossing. Using standard terminology in
LZ transitions \citep{key-GomezLlorenteLZ,key-NoriStuec}, one can
say that, when the ramping process starts, the adiabatic states approximately
match their diabatic counterparts. Notice that it is a three-level
LZ model that corresponds to $F=1$. As the ramping goes on, the preparation
proceeds through a slow decrease of $\left|\Delta\right|$ till $\Delta=0$
is reached. At that point, the target dressed state is attained. \medskip{}
\medskip{}

\includegraphics[scale=0.6]{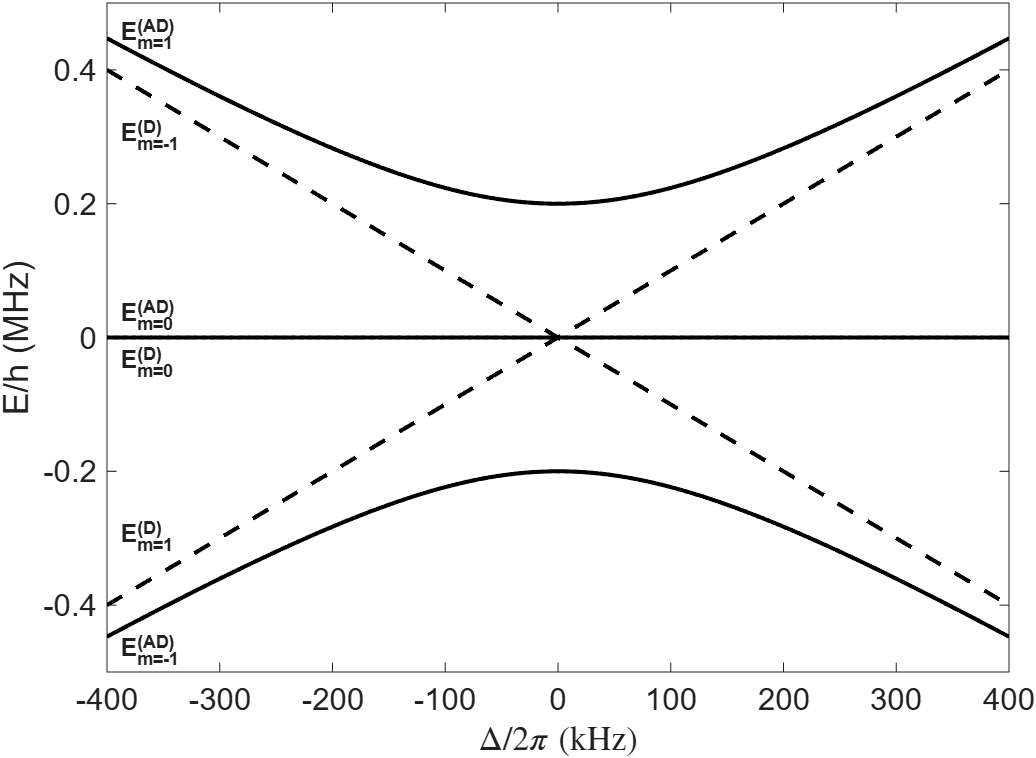}

\begin{figure}[H]
\caption{Bare energy levels (dashed lines) and dressed counterparts (continuous
lines) of the three-level LZ model corresponding to the Hamiltonian
in Eq. (\ref{eq:RotatedBasicHamiltonian}) with $F=1$, $\Omega_{d}/(2\pi)=0.2\,\textrm{(MHz)}$.
For $m=0$, the bare and the dressed energy levels are equal (horizontal
line)}

\end{figure}

In the absence of noise, the adiabatic-passage scheme is guaranteed
to work: for a sufficiently slow ramping, the system follows the initially
prepared adiabatic eigenstate. This is shown applying the results
of the analysis presented in Ref. \citep{key-LZ3Zenesini} for an
analogous (noiseless) three-level LZ model. Specifically, we depart
from the left part of the diagram with $\left|\Delta\right|\gg\Omega_{d}$
and with the system in the diabatic state $\left|F=1,m=-1\right\rangle ^{(D)}$,
which approximately matches the adiabatic state $\left|F=1,m=1\right\rangle ^{(AD)}$.
Then, it is found that, if the ramp is not interrupted, (i.e., if
it is continued after $\Delta=0$ till reaching $\left|\Delta\right|\gg\Omega_{d}$
in the right), the probability of remaining in the associated adiabatic
state is given by 

\begin{equation}
P_{1}=(1-p)^{2},\label{eq:LZ3}
\end{equation}
 where 
\begin{equation}
p=\exp(-2\pi\frac{\left|\Omega_{d}\right|^{2}}{2\hbar\left|\dot{\Delta}\right|}),\label{eq:parameterLZ}
\end{equation}
 with $\dot{\Delta}=\dot{\omega}_{0}$ being the rate of detuning
variation. In our system, the ramp ends at $\Delta=0$; still, given
the LZ characteristics of the process, the expression given by Eq.
(\ref{eq:LZ3}) can be soundly considered to approximate the magnitude
of the probability of permanence in the considered adiabatic state.
Therefore, from Eqs (\ref{eq:LZ3}) and (\ref{eq:parameterLZ}), it
follows that, for a sufficiently slow ramp, $\frac{\left|\Omega_{d}\right|^{2}}{2\hbar\left|\dot{\Delta}\right|}\gg1$,
the probability of a transition to other adiabatic state is negligible. 

Now, to assure the applicability of the scheme in a realistic (noisy)
scenario, the effect of the fluctuations on the transitions must be
evaluated. This issue is elucidated through the following general
considerations: 

i) From the analysis of the transformed Hamiltonian, {[}see Eq. (\ref{eq:preparationHamiltonian}){]},
it is apparent that the diagonal term, $\chi_{c}(t)\hat{F}_{z}$,
merely introduces a stochastic phase in the evolution of each of the
dressed states. That term does not compromise the intended preparation.
In contrast, one cannot \emph{a priori} discard the potential transfer
of population that can result from the stochastic transverse component
in Eq. (\ref{eq:preparationHamiltonian}). From the following arguments,
that possibility is discarded.

ii) The separation between adiabatic energies is sufficiently large
for evaluating the effect of the random contribution, $\chi_{d}(t)\hat{F}_{x}+\chi_{b}(t)\hat{F}_{y}$,
using a perturbative scheme. This is inferred from the working conditions:
even at the smallest adiabatic energy separation, reached at $\Delta=0$,
the noise effect on the population transfer, as considered in the
former section, can be regarded as a perturbation. Moreover, as previously
shown, the efficiency of the stochastic terms to induce population
transfers is determined by  the magnitude of the associated spectral
densities at the transition frequency. That frequency, given by $\sqrt{\Delta(t)^{2}+\Omega^{2}_{d}(t)}$,
 varies from $\left|\Delta\right|\gg\Omega_{d}$ (at the beginning
of the ramp), to $\Omega_{d}$ (at the end of the process, where $\Delta=0$). 

iii) Along with the above argument, we must take into account that
the spectra of $\chi_{d}(t)$ and $\chi_{b}(t)$ are built up from
those of the original random terms $\delta\omega_{0}(t)$, $\eta_{x}(t)$,
and $\eta_{y}(t)$. Here, some findings of the previous sections must
be recalled. The assumed requirements for the efficiency of the CDD
technique to curb diagonal-noise effects include a significant reduction
of the spectral density of $\delta\omega_{0}(t)$ at the effective
frequency $\tilde{\Omega}_{d}$. Additionally, because of the shifts
in $\pm\omega_{d}$ present in Eqs. (\ref{eq:SpectrumXia}) and (\ref{eq:SpectrumXib}),
an important reduction in the spectral densities of $\chi_{a}(t)$
(and, in turn, in those of $\chi_{d}(t)$) and of $\chi_{b}(t)$)
with respect to the original noise input takes place since the RWA
restriction $\omega_{d}\gg\Omega_{d}$ must be fulfilled. That restriction
also mitigates the effects of the cross-correlations. 

iv) Combining the arguments of the above points, it is concluded that,
in the whole ramping process, because of the reduced values of the
spectral densities of $\chi_{d}(t)$ and $\chi_{b}(t)$ at the effective
transition frequencies, the occurrence of significant changes in the
state populations can be ruled out. The applicability of this analysis
is illustrated by the dressed-state preparation realized in the experimental
setup of Ref. \citep{key-SpielmanClockTransitions}. 

v) The above considerations are not applicable to the case where all
the fluctuations have white-noise character, which has been excluded
from the analysis. In Fig. 5, we depict the spectral densities of
$\delta\omega_{0}(t)$, $\chi_{a}(t)$, and $\chi_{d}(t)$ for different
correlation times. To simplify the representation, we have not included
$\chi_{c}(t)$, which enters the Hamiltonian longitudinally, and,
as discussed, has no effect on the dressed-state preparation. We focus
on the transverse term $\chi_{d}(t)$, defined in terms of $\delta\omega_{0}(t)$
and $\chi_{a}(t)$. {[}The term $\chi_{b}(t)$, also transverse, has
the same spectral density as $\chi_{a}(t)${]}. It is observed that,
as the correlation times decrease, the concentration of the effective-noise
spectrum around $\omega_{d}$ is attenuated, and, consequently, the
probability for a noise-induced transition between adiabatic states
can become relevant.\medskip{}
\medskip{}

\textcolor{red}{\includegraphics[clip,scale=0.6]{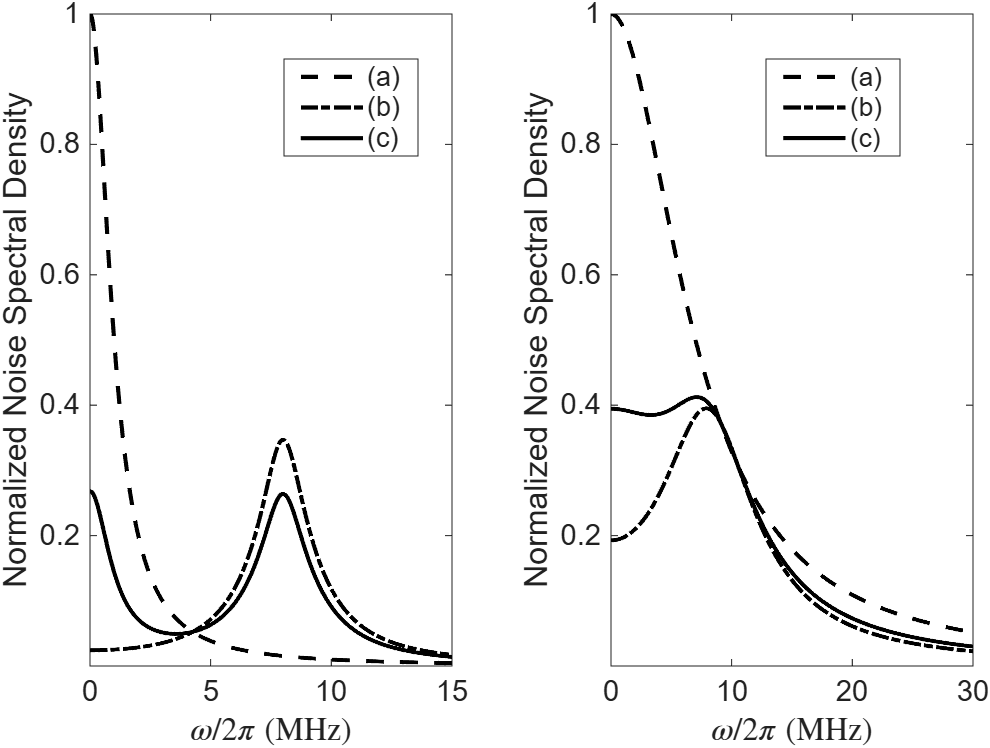}}

\begin{figure}[H]
\caption{Spectral densities of the fluctuations $\delta\omega_{0}(t)$ (a),
$\chi_{a}(t)$ (b), and $\chi_{d}(t)$ (c), normalized to the maximum
of $S(\delta\omega_{0};\omega)$. In the left, the input fluctuations
correspond to $\left\langle \eta^{2}_{x}\right\rangle =150\textrm{\,(a.u.)}$,
$\alpha_{\eta_{x}}/2\pi=1\,\textrm{(MHz)}$, $\left\langle \eta^{2}_{y}\right\rangle =250\textrm{(a.u.)}$,
$\alpha_{\eta_{y}}/2\pi=2\,\textrm{(MHz)}$, $\left\langle \delta\omega^{2}_{0}\right\rangle =200\textrm{\,(a.u.)}$,
and, $\alpha_{\delta\omega_{0}}=1\,\textrm{(MHz)}$. In the right,
$\left\langle \eta^{2}_{x}\right\rangle =150\textrm{\,(a.u.)}$, $\alpha_{\eta_{x}}/2\pi=4.5\,\textrm{(MHz)}$,
$\left\langle \eta^{2}_{y}\right\rangle =50\textrm{\,(a.u.)}$, $\alpha_{\eta_{y}}/2\pi=6\,\textrm{(MHz)}$,
$\left\langle \delta\omega^{2}_{0}\right\rangle =200\textrm{\,(a.u.)}$,
and, $\alpha_{\delta\omega_{0}}=7\,\textrm{(MHz)}$. In both, left
and right figures, $\omega_{d}/2\pi=8\,\textrm{(MHz)}$ and $\theta=\pi/3$. }
\end{figure}

\section{Concluding remarks}

The study has shown that the CDD method, originally intended to mitigate
dephasing due to longitudinal fluctuations, can reduce also the deleterious
effects on coherence of transverse noise potentially present in the
practical setups. Indeed, a global understanding of the efficiency
of CDD to curb, both, longitudinal and transverse fluctuations has
been given. Crucial to our analysis has been the description of the
fluctuations in the dressed state-basis associated to  the CDD field
of control. In that representation, the random terms acquire forms
qualitatively different from those of the original stochastic inputs.
In particular, the effective spectral densities, which become concentrated
on a range centered on the driving-field frequency, are significantly
reduced at the (much smaller) fundamental frequencies relevant to
the dynamics. As a consequence, the dephasing rooted in the longitudinal
noise and the population transfers resulting from transverse fluctuations
can be substantially reduced. For both, diagonal and transverse fluctuations,
the efficiency of the CDD technique declines as the noise correlation
time decreases. Whereas, in the white-noise limit, the choice of the
driving parameters has no applicability as a strategy to extend the
coherence time, in the opposite limit, i.e., when a static scenario
for the fluctuations can be assumed, a high efficiency can be achieved. 

The generation of the dressed states via adiabatic passage has been
found to be robust against the presence of fluctuations with characteristics
far from the white-noise limit. Actually, no appreciable values of
the probability for noise-induced transitions between adiabatic states
are found since the CDD method induces the reduction of the spectral
densities of the effective stochastic terms at the frequencies of
the potential transitions. As a consequence, the presence of noise
with narrow spectra does not significantly affect the adiabatic preparation
of the dressed states.

Our results have uncovered the role of noise anisotropy in the appearance
of rapidly oscillating terms in the system response to the fluctuations.
In the considered range of parameters, those oscillations have been
found to be secondary to the deterministic corrections to the RWA.
The analogy with the behavior observed in Ref. \citep{key-AnisotropicTransverseNoise}
has been traced. Indeed, the platform of \citep{key-AnisotropicTransverseNoise}
can be considered to provide an appropriate test bench for our predictions
on the efficiency of CDD in transverse-noise settings.

The use of our analytical results in the design of strategies to extend
the coherence times can be expected. However, one must take into account
that, to guarantee the applicability of the approach, the variation
of the control parameters must be restricted to specific ranges. First,
the driving frequency must be close to the qubit frequency. Second,
the field amplitude must be small enough for the RWA to be valid. 

The data that support the findings of this article are openly available
\citep{key-data}.

\medskip{}
\medskip{}
\medskip{}

\medskip{}
\medskip{}
\medskip{}
\medskip{}

\section*{Appendix A: Dynamics beyond the Rotating Wave Approximation}

The analysis of the effects of static noise has been carried out without
resorting to the RWA, previously applied in the characterization of
the \emph{deterministic} dynamics. Indeed, we have analytically uncovered
the existence of oscillations with frequency $2\omega_{d}$ associated
to anisotropy in the noisy input. It is apparent that, in order to
evaluate the actual relevance of those oscillations, we must set up
a framework where both, deterministic and stochastic components, are
studied at the same order of approximation. Accordingly, in this Appendix,
we present the description of the noiseless dynamics beyond the RWA. 

The application of the unitary transformation given by Eq. (\ref{eq:FirstUnitary})
to the Hamiltonian in Eq. (\ref{eq:BasicHamiltonian}) in the absence
of the random contributions leads to 

\begin{eqnarray}
\hat{H} & = & \Delta\hat{F}_{z}+\Omega_{d}\left[1+\cos(2\omega_{d}t)\right]\hat{F}_{x}-\Omega_{d}\sin(2\omega_{d}t)\hat{F}_{y},\label{eq:NRWA}
\end{eqnarray}
 where we have retained the terms oscillating with frequency $2\omega_{d}$.
Now, through the rotation characterized by $\hat{U}_{2}(t)$ {[}see
Eq. (\ref{eq:Unitary2}){]}, the Hamiltonian, in the case of zero
detuning, is cast into the form 

\begin{eqnarray}
\hat{H} & = & \Omega_{d}\left[1+\cos(2\omega_{d}t)\right]\hat{F}_{z}-\Omega_{d}\sin(2\omega_{d}t)\hat{F}_{y},\label{eq:NRWArotated}
\end{eqnarray}
 which, for $\omega_{d}\gg\Omega_{d}$, can be regarded as the sum
of a zero-order term 

\[
\hat{H}_{0}(t)=\Omega_{d}\left[1+\cos(2\omega_{d}t)\right]\hat{F}_{z},
\]
 and a time-dependent perturbation 
\[
\hat{W}(t)=-\Omega_{d}\sin(2\omega_{d}t)\hat{F}_{y}.
\]

Working in the associated perturbative scheme, our procedure starts
with the characterization of the dynamics governed by $H_{0}(t)$.
The evolution of the density matrix in the rotating frame $\hat{U}_{3}(t)=e^{-i\Omega_{d}t\hat{F}_{z}/\hbar}$
is shown to be given by 

\begin{equation}
\rho_{m,m^{\prime}}(t)=\rho_{m,m^{\prime}}(0)e^{i(m-m^{\prime})\frac{\Omega_{d}}{2\omega_{d}}\sin(2\omega_{d}t)}.\label{eq:coherencesNRWA}
\end{equation}
 Note that the magnitude of the amplitude of the exponent oscillation
is given by $\frac{\Omega_{d}}{\omega_{d}}$. {[}The comparison with
the stochastic analogue given by Eq. (\ref{eq:StaticDephasing}) was
presented in Sec (III-A-2){]}. Convenient to our analysis is the use
of the Jacobi-Anger expansion, given by \citep{key-Grad}, 
\begin{equation}
\exp(iz\sin\varphi)=J_{0}(z)+2\sum^{\infty}_{k=1}J_{2k}(z)\cos(2k\varphi)+2i\sum^{\infty}_{k=0}J_{2k+1}(z)\sin[2(k+1)\varphi],\label{eq:BesselExpansion}
\end{equation}
 where $J_{n}(z)$ are the ordinary Bessel functions. Taking $z=(m-m^{\prime})\frac{\Omega_{d}}{2\omega_{d}}$
and $\varphi=2\omega_{d}t$, and introducing the expansion into Eq.
(\ref{eq:coherencesNRWA}), the coherences are found to present a
nontrivial evolution for a generic value of the argument $z$: different
 terms oscillating with frequencies multiples of $2\omega_{d}$ can
contribute to the expansion. As the magnitude of $z$ decreases, the
evolution becomes significantly simplified: for $z\ll1$, the magnitude
of the different oscillating terms, characterized by $J_{n}(z)$,
$n\neq0$, decays as $n$ grows. Hence, we can make the approximation 

\begin{equation}
\rho_{m,m^{\prime}}(t)\simeq\rho_{m,m^{\prime}}(0)\left[J_{0}(z)+2iJ_{1}(z)\sin(2\omega_{d}t)\right],\label{eq:coherencesNRWA-1}
\end{equation}
which reflects that the dominant oscillating contribution to the coherences
can be traced to the (fundamental) frequency $2\omega_{d}$. In the
limit $z\rightarrow0$, the magnitude of the oscillating terms becomes
negligible ($J_{1}(z)\rightarrow0$, $J_{0}(z)\rightarrow1$), and,
we consistently recovered the result obtained in the RWA  framework.

As a second step in our procedure, we apply now time-dependent perturbation
theory to describe the population transfer induced by $\hat{W}(t)$
between the eigenstates of $\hat{H}_{0}(t)$. Specifically, neglecting
the second-order effect associated to the time modulation induced
by the term $\frac{\Omega_{d}}{2\omega_{d}}\sin(2\omega_{d}t$), rooted
in the time dependence of the eigenvalues of $\hat{H}_{0}(t)$, one
obtains 

\begin{eqnarray}
P_{m,m^{\prime}}(t) & = & \frac{1}{\hbar^{2}}\left|\int^{t}_{0}dt^{\prime}W_{m,m^{\prime}}(t^{\prime})e^{i(m-m^{\prime})\Omega_{d}t^{\prime}}\right|^{2}\nonumber \\
 & \simeq & \mathcal{F}^{(y)}_{m,m^{\prime}}\Omega^{2}_{d}\left[\frac{\sin^{2}(\frac{\Omega_{+2}t}{2})}{\Omega^{2}_{+2}}+\frac{\sin^{2}(\frac{\Omega_{-2}t}{2})}{\Omega^{2}_{-2}}+\frac{\cos(\Omega_{+2}t)+\cos(\Omega_{-2}t)-2\cos^{2}(2\omega_{d}t)}{2\Omega_{+2}\Omega_{-2}}\right]\label{eq:PopTransNRWA}
\end{eqnarray}
 where $\Omega_{+2}=\tilde{\Omega}_{d}+2\omega_{d}$, and $\Omega_{-2}=\tilde{\Omega}_{d}-2\omega_{d}$.
Hence, oscillations with frequency $2\omega_{d}$ are apparent in
the transfer probability. Observe that the magnitude of the oscillation
amplitude can be approximated by $\mathcal{F}^{(y)}_{m,m^{\prime}}\Omega^{2}_{d}/\omega^{2}_{d}$.
{[}The comparison with the stochastic counterpart given by Eq. (\ref{eq:StaticLimitTransfer})
was presented in Sec. (III-B-2){]}. The connection with the RWA is
consistently established: the transfer becomes negligible as the quotient
$\frac{\Omega_{d}}{\omega_{d}}$ decreases and the system enters the
range of applicability of the RWA. 

\section*{Appendix B: Dressed-state population transfer as a function of the
noise spectral densities}

Here, we present some details of the derivation of Eq. (\ref{eq:AveragedPopulTransfer}).
To evaluate the dressed-state population transfer, we depart from
Eq. (\ref{eq:GeneralAverPopTransfer}), and, taking into account that
$\Omega_{d}t\gg\zeta_{a}(t)$, neglect the random shift in the frequency.
Then, we calculate each of the two contributions. We start by the
term 

\begin{eqnarray}
\mathcal{F}^{(x)}_{m,m^{\prime}}\left\langle \left|\int^{t}_{0}dt^{\prime}\delta\omega_{0}(t^{\prime})e^{i(m-m^{\prime})\Omega_{d}t^{\prime}}\right|^{2}\right\rangle  & = & \mathcal{F}^{(x)}_{m,m^{\prime}}\left\langle \int^{t}_{0}d\tau\delta\omega_{0}(\tau)e^{i\tilde{\Omega}_{d}\tau}\times\int^{t}_{0}d\tau^{\prime}\delta\omega_{0}(\tau^{\prime})e^{-i\tilde{\Omega}_{d}\tau^{\prime}}\right\rangle 
\end{eqnarray}
 which, leaving out the factor $\mathcal{F}^{(x)}_{m,m^{\prime}}$,
can be simply rewritten as 
\begin{equation}
\int^{t}_{0}d\tau\int^{t}_{0}d\tau^{\prime}\left\langle \delta\omega_{0}(\tau)\delta\omega_{0}(\tau^{\prime})\right\rangle e^{-i\tilde{\Omega}_{d}(\tau^{\prime}-\tau)}=\int^{t}_{0}d\tau\int^{t}_{0}d\tau^{\prime}G(\delta\omega_{0};\tau^{\prime}-\tau)e^{-i\tilde{\Omega}_{d}(\tau^{\prime}-\tau)}.
\end{equation}
 Now, making an appropriate change of variables \citep{key-Stratonovich},
we can express it as 
\begin{equation}
\int^{t}_{-t}d\tau(t-\left|\tau\right|)G(\delta\omega_{0};\tau)e^{-i\tilde{\Omega}_{d}\tau}.
\end{equation}
 Additionally, using the Wiener-Khinchin theorem and evaluating the
integral in $\tau$, we cast that expression into 

\begin{equation}
\int^{\infty}_{-\infty}d\omega S(\delta\omega_{0};\omega)\left(\frac{\sin\left[(\omega-\tilde{\Omega}_{d})t/2\right]}{(\omega-\tilde{\Omega}_{d})/2}\right)^{2}.
\end{equation}
 Finally, we take into account that, at times much larger than the
inverse of the spectral width ($t\gg\tau_{c}$), it is possible to
approximate the integrand in the form \citep{key-GomezLlorenteNoiseSpectrum,key-Paladino1/fNoiseReview}

\begin{equation}
S(\delta\omega_{0};\omega)\left(\frac{\sin\left[(\omega-\tilde{\Omega}_{d})t/2\right]}{(\omega-\tilde{\Omega}_{d})/2}\right)^{2}\sim2\pi tS(\delta\omega_{0};\omega)\delta(\omega-\tilde{\Omega}_{d}).
\end{equation}
 Actually, $\left(\frac{\sin\left[(\omega-\tilde{\Omega}_{d})t/2\right]}{(\omega-\tilde{\Omega}_{d})/2}\right)^{2}$is
a highly-peaked function centered on $\omega=\tilde{\Omega}_{d}$,
whereas $S(\delta\omega_{0};\omega)$ varies smoothly. Accordingly,
we find 

\begin{eqnarray}
\mathcal{F}^{(x)}_{m,m^{\prime}}\left\langle \left|\int^{t}_{0}dt^{\prime}\delta\omega_{0}(t^{\prime})e^{i(m-m^{\prime})\Omega_{d}t^{\prime}}\right|^{2}\right\rangle  & \simeq & 2\pi t\mathcal{F}^{(x)}_{m,m^{\prime}}S(\delta\omega_{0};\tilde{\Omega}_{d})
\end{eqnarray}

Following a similar procedure, the second contribution in Eq. (\ref{eq:GeneralAverPopTransfer})
is evaluated to obtain 

\begin{equation}
\mathcal{F}^{(y)}_{m,m^{\prime}}\left\langle \left|\int^{t}_{0}dt^{\prime}\chi_{b}(t^{\prime})e^{i(m-m^{\prime})\Omega_{d}t^{\prime}}\right|^{2}\right\rangle \simeq2\pi t\mathcal{F}^{(y)}_{m,m^{\prime}}S(\chi_{b};\tilde{\Omega}_{d}).\label{eq:SecondContrib}
\end{equation}

\section*{Appendix C: The effect of amplitude noise}

In this appendix, we generalize our description to deal with the effect
of amplitude noise. Fluctuations in the amplitude of the driving field
are incorporated into our theoretical approach by replacing $\Omega_{d}$
in Eq. (\ref{eq:BasicHamiltonian}) by $\Omega_{d}+\delta\Omega_{d}(t)$,
where $\delta\Omega_{d}(t)$ represents a stochastic variable, which,
without loss of generality, can be assumed to have zero mean-value.
Note that amplitude noise is actually transverse noise; the essential
difference with the formerly considered (transverse) fluctuations
$\eta_{x}(t)$ and $\eta_{y}(t)$ is the driving factor that multiplies
$\delta\Omega_{d}(t)$. We stress that no concatenation scheme is
implemented here; our objective is to evaluate if the basic scheme,
which has been shown to be robust against undriven transverse noise,
is also useful to tackle fluctuations in the amplitude. 

Paralleling the previous procedure, we apply the sequence of unitary
transformations $\hat{U}_{1}(t)$, $\hat{U}_{2}(t)$, and $\hat{U}_{3}(t)$
to the modified Hamiltonian. Subsequently, we set up a perturbative
scheme where the zero-order Hamiltonian is given now by 

\begin{equation}
\hat{H}_{0}(t)=[\delta\Omega_{d}(t)+\chi_{a}(t)]\hat{F}_{z}.\label{eq:AmplitudeNoiseHcero}
\end{equation}
 Note that, in the considered rotating frame defined by $\hat{U}_{1}(t)$,
amplitude noise becomes undriven; in contrast, $\eta_{x}(t)$ and
$\eta_{y}(t)$ are incorporated into the effective driven term $\chi_{a}(t)$.
From $\hat{H}_{0}(t)$, the time evolution for a stochastic trajectory
is written as 

\begin{equation}
\left|\psi(t)\right\rangle =e^{-i\tilde{\zeta_{a}}(t)\hat{F}_{z}/\hbar}\left|\psi(0)\right\rangle ,
\end{equation}
where $\tilde{\zeta_{a}}(t)$ is the non-stationary random variable
defined by

\begin{equation}
\tilde{\zeta_{a}}(t)=\int^{t}_{0}\left[\delta\Omega_{d}(t^{\prime})+\chi_{a}(t^{\prime})\right]dt^{\prime}.\label{eq:AmplitudeNFirstStochVariable}
\end{equation}
 In turn, the evolution of the coherences, obtained by averaging over
stochastic trajectories, is shown to be given by 
\begin{equation}
\left\langle \rho_{m,m^{\prime}}(t)\right\rangle =\rho_{m,m^{\prime}}(0)\left\langle e^{i(m-m^{\prime})\tilde{\zeta_{a}}(t)}\right\rangle ,\label{eq:AMplitudeNcoherences}
\end{equation}
Once the properties of amplitude noise in the specific considered
setting are identified, we can proceed to carry out the statistical
average. This step is particularly direct if $\delta\Omega_{d}(t)$
can be considered to have static Gaussian character and to be uncorrelated
with $\chi_{a}(t).$ Indeed, in that case, the coherences are shown
to evolve as 

\begin{eqnarray}
\left\langle \rho_{m,m^{\prime}}(t)\right\rangle  & = & \rho_{m,m^{\prime}}(0)\left\langle e^{i(m-m^{\prime})\delta\Omega_{d}(0)t}\times e^{i(m-m^{\prime})\zeta_{a}(t)}\right\rangle \label{eq:AmplitudeNdephasing}\\
 & = & \rho_{m,m^{\prime}}(0)e^{-\frac{1}{2}(m-m^{\prime})^{2}\left\langle \delta\Omega^{2}_{d}\right\rangle t^{2}}\left\langle e^{i(m-m^{\prime})\zeta_{a}(t)}\right\rangle ,
\end{eqnarray}
where $\left\langle \delta\Omega^{2}_{d}\right\rangle $ represents
the variance of the fluctuations. {[}Observe that, because of the
static-noise characteristics, we have taken $\delta\Omega_{d}(t)=\delta\Omega_{d}(0)\equiv\delta\Omega_{d}${]}.
Hence, a more complex form of the coherence decay becomes apparent:
the static amplitude noise introduces a factor of Gaussian exponential
decay in the previously analyzed dephasing. The decoherence time in
this modified scenario depends on the magnitude of the variances of
the different noises involved {[}$\delta\Omega_{d}$, $\eta_{x}(t)$
and $\eta_{y}(t)${]}, and on the value of the driving frequency.
It is important to emphasize that, whereas, as previously shown, the
dephasing effect of $\chi_{a}(t)$ can be significantly mitigated
by working with a sufficiently large $\omega_{d}$, that strategy
does not work for amplitude noise, which, in the considered rotating
frame, is not modulated by terms oscillating with $\omega_{d}$. Hence,
the presence of $\delta\Omega_{d}(t)$ can significantly reduce the
decoherence time. As an illustrative example, let us evaluate how
dephasing in the system characterized by the different sets of parameters
used in Fig. 3, which was shown to correspond to simple exponential
decay, is modified by the presence of amplitude noise. It is found
that, for (static Gaussian) amplitude fluctuations with a standard
deviation $\sqrt{\left\langle \delta\Omega^{2}_{d}\right\rangle }$
in the range ($0.01-0.1)\Omega_{d}$, ($\Omega_{d}\sim0.1\omega_{d}$),
it is indeed the amplitude noise that determines the magnitude of
the dephasing time in all the situations illustrated in Fig. 3. Specifically,
for $\omega_{d}/2\pi=10\,MHz$, we find that $T_{2}$ is in the range
$\sim(0.2-2)\,\mu s$. 

\bigskip{}
\bigskip{}

\end{document}